\documentclass{elsart}
\usepackage{amssymb}
\usepackage{amsmath}
\usepackage{graphicx}
\newcommand{\ie}{i.e.}
\newcommand{\nn}{n.n.}
\newcommand{\nnn}{n.n.n.}

\begin{document}
\runauthor{Derzhko and J{\c{e}}drzejewski}
\begin{frontmatter}

\title{Formation of charge-stripe phases in a system
of spinless fermions or hardcore bosons}

\author[Wroclaw]{Volodymyr Derzhko} and
\author[Wroclaw,Lodz]{Janusz J{\c{e}}drzejewski\thanksref{email}}

\address[Wroclaw]{Institute of Theoretical Physics,
University of Wroc{\l}aw, pl. Maksa Borna 9, 50--204 Wroc{\l}aw,
Poland}
\address[Lodz]{Department of Theoretical Physics, University of
{\L}{\'{o}}d{\'{z}}, ul. Pomorska 149/153, 90--236 {\L}\'{o}d\'{z},
Poland}

\thanks[email]{Corresponding author: J. J{\c{e}}drzejewski, phone:
+48 71 3759415, fax: +48 71 3214454, e-mail: jjed@ift.uni.wroc.pl}

\begin{abstract}
We consider two strongly correlated two-component quantum systems,
consisting of quantum mobile particles and classical immobile
particles. The both systems are described by Falicov--Kimball-like
Hamiltonians on a square lattice, extended by direct short-range
interactions between the immobile particles. In the first system the
mobile particles are spinless fermions while in the second one they
are hardcore bosons. We construct rigorously ground-state phase
diagrams of the both systems in the strong-coupling regime and at
half-filling. Two main conclusions are drawn. Firstly,
short-range interactions in quantum gases are sufficient for the
appearance of charge stripe-ordered phases. By varying the
intensity of a direct nearest-neighbor interaction between the
immobile particles, the both systems can be driven from a
phase-separated state (the segregated phase) to a crystalline
state (the chessboard phase) and these transitions occur
necessarily via charge-stripe phases: via a diagonal striped phase
in the case of fermions and via vertical (horizontal) striped
phases in the case of hardcore bosons. Secondly, the phase
diagrams of the two systems (mobile fermions or mobile hardcore
bosons) are definitely different. However, if the strongest
effective interaction in the fermionic case gets frustrated
gently, then the phase diagram becomes similar to that of the
bosonic case.

\end{abstract}

\end{frontmatter}

\section{Introduction}

In the passing decade, some specific phases, having
quasi-one-dimensional structure, the so called striped phases, have
been a highly debated subject in condensed-matter physics.
Apparently, a broad interest in such phases was initiated by reports
presenting experimental evidence for the existence of charge strips
in doped layered perovskites, some of which constitute materials
exhibiting high-temperature superconductivity
\cite{tranquada1,tranquada2}.

However, striped phases had been observed much earlier, for instance
on metallic surfaces covered with an adsorbate, see \cite{kern} and
references therein. Theoretical descriptions of these phenomena
involved various kinds of lattice gas models \cite{sasaki}. Many
other instances of experimental observations of stripe-ordered
phases are listed in \cite{malescio}, where Monte-Carlo studies of
formation of striped phases in a continuous gas with hardcore and
short-range repulsive interactions are reported.

Quite interestingly, theoretical studies of striped phases in
systems of strongly correlated electrons have preceded
experimental observations of such phases \cite{poilblanc,zaanen}.
But only after those observations, the discussion became much more
vigorous, and the nature of stripe-ordered phases started to
attract attention of numerous researchers. In the context of the
Hubbard model, a comprehensive review of the problem can be found
in \cite{oles}. The existence of the same kind of striped phases
was investigated also in the $t$--$J$ model \cite{white1,white2}.
A bird's eye view on the problem of stripe-ordered phases in
high-temperature superconductors, but emphasizing its general
relevance for contemporary condensed-matter physics can be found
in \cite{emery}. The general relevance of striped phases is
underlined also in \cite{zhang1}, where they are viewed as an
emergent phenomenon resulting from collective motions of
microscopic particles, somewhat analogous to quasi-particles like
phonons. Striped phases have been also studied in the framework of
quantum-spin models, like $XY$ model for instance \cite{sandvik}.

The Hubbard or $t$--$J$-like models belong to the most realistic
models, in the framework of which the problem of striped-ordered
phases in doped layered perovskites can be investigated. In the both
models, the spin and the charge degrees of freedom are taken into
account, and it is believed that it is the competition between these
degrees of freedom that is decisive for the formation of striped
phases. The question of formation of striped phases is closely
related to the question of relative stability of these phases
against mixtures of an electron-reach phase and a hole-reach phase.
Due to the tiny energy differences between both phases, the results
obtained by means of approximate methods, which introduce
hardly-controllable errors, are disparate. And this can be said
despite, as emphasized in \cite{oles}, a spectacular consensus of
results concerning half-filled stripes, obtained by means of various
approximate methods. Therefore, as pointed out in
\cite{zhang1,lemanski1} further careful studies are necessary to
settle the problem of formation of stripe-ordered phases. One of
possible ways of attacking this problem is to formulate analog
problems in less realistic but simpler models, where some control
over the results obtained by means of various methods of many-body
physics can be gained. This leads hopefully to a deeper insight into
the, mentioned above, stability problem.

Such an approach has been adopted by many researchers, see for
instance \cite{zhang1,buhler,lemanski1,lemanski2,zhang2}. Buhler et
al \cite{buhler} have found, by means of Monte Carlo simulations,
that upon hole doping antiferromagnetic spin domains and charge
stripes, whose properties are in very good agreement with
experiments, appear in a spin-fermion model for cuprates. Using the
so called restricted phase diagrams, the stability problem of
charge-stripe phases has been studied in the spinless
Falicov--Kimball model by Lemanski et al \cite{lemanski1,lemanski2}.
In their study the formation of charge stripe-ordered phases can be
looked upon as a way the system interpolates between a periodic
charge-density wave phase (the chessboard phase) and the segregated
phase (a mixture of completely filled and completely empty phases),
as the degree of doping varies. A considerable reduction of the
Hilbert space dimension in a spinless fermion model with infinite
nearest-neighbor repulsion (as compared to a Hubbard model) has been
exploited by Zhang and Henley \cite{zhang1,zhang2} to study
carefully, by means of an exact diagonalization technique, the
formation of charge stripe-ordered phases upon doping. They
addressed also an interesting question of the role of quantum
statistics in the problem of striped phases, by replacing fermion
particles with hardcore boson ones.

Our work has been inspired mainly by the recent studies of
Lema{\'{n}}ski et al \cite{lemanski1,lemanski2} and by Zhang and
Henley \cite{zhang1,zhang2}. To investigate the key question,
whether charge stripe-ordered phases are stable compared to a
phase-separated state, we study rigorously the strong-coupling
limit of an extended spinless Falicov--Kimball model on a square
lattice, with mobile particles being spinless fermions or hardcore
bosons. The usual spinless Falicov--Kimball  Hamiltonian (such as
that studied in \cite{lieb1}) has been augmented by a direct,
Ising-like, interaction between the immobile particles.

To explain the role played by the extra interaction we have to make
a few remarks. Firstly, for technical reasons, our analysis is
restricted to the case of half-filling. Secondly, it is now well
established (proven) \cite{lieb1,lieb2} that, in the fermion
spinless Falicov--Kimball model the phase-separated state, the
segregated phase, is stable only off the half-filling. Thus, an
additional interaction is needed to stabilize the segregated phase
at half-filling. Thirdly, in the regime of singly occupied sites and
for strong-coupling, the strongest effective interaction in the
Hubbard model is the Heisenberg antiferromagnetic interaction.
Doping of holes can be seen as means to weaken the strong tendency
toward antiferromagnetic ordering and to make possible a phase
separated state with hole-reach and electron-reach regions. Under
similar conditions, in the spinless Falicov--Kimball model an
analogous role is played by the Ising-like nearest-neighbor ({\nn})
repulsive interaction between the immobile particles that favors a
chessboard-like ordering. The effect of weakening of the strong
tendency toward chessboard ordering can be achieved by an extra
Ising-like \nn{} interaction that compensates the strongest repulsive
interaction, and consequently permits the system to reach a
phase-separated state, the segregated phase. On varying, in a
suitable interval of values, the corresponding interaction constant,
which is our control parameter, we can study how the both systems
``evolve'' from the crystalline chessboard phase to the segregated
one.

We prove that by varying the interaction constant of the direct \nn{}
interaction between the immobile particles, the both systems
(fermion and boson) can be driven, via a sequence of phase
transitions, from a crystalline state (the chessboard phase) to a
phase-separated state (the segregated phase). Moreover, these
transitions occur necessarily via charge-stripe phases: via a
diagonal striped phase in the case of fermions and via vertical
(horizontal) striped phases in the case of hardcore bosons. Thus,
short-range interactions in quantum gases are sufficient for the
appearance of charge stripe-ordered phases.

In our studies, we include also a subsidiary direct interaction
between the immobile particles, an Ising-like next nearest-neighbor
({\nnn}) interaction, much weaker than the \nn{} interaction. This
interaction can reinforce or frustrate the \nn{} interaction,
depending on the sign of its interaction constant. It appears that
on varying the control parameter (the strength of \nn{}
interactions), the systems (fermion or boson) characterized by
different values of \nnn{} interactions, may undergo different
sequences of transitions between the chessboard phase and the
segregated one. We find, in particular, that if we set appropriate
values of the \nnn{} interaction in the fermion system and in the
boson one, then the both systems settle in the same phases, for
typical values of the control parameter. On the other hand, the
\nnn{} interaction  enables us to demonstrate that the ground-state
phase diagrams in the cases of mobile fermions and mobile hardcore
bosons are definitely different.

As a byproduct of our considerations, we obtain an Ising-like
short-range Hamiltonian, with pair interactions at distance of two
lattice constants and  four-site (plaquette) interactions, whose
ground states consist of stripe-like configurations or
chessboard-like configurations, depending on the sign of the unique
coupling constant. These ground states are stable with respect to
perturbations by \nn{} and \nnn{} interactions.

The paper is organized as follows. In the next section we present
the considered models and general properties of the corresponding
Hamiltonians. Then, in section 3 we construct phase diagrams due to
truncated effective interactions. After that, in section 4, we
analyze  the phase diagrams due to complete interactions (without
truncation) of the both quantum systems studied here and formulate
our conclusions. Finally, in section 5, we summarize our results.

\section{The model and its basic properties}

The model to be studied is a simplified version of the one band,
spin $1/2$ Hubbard model, known as the static approximation (one
sort of electrons hops while the other sort is immobile), augmented
by a direct Ising-like interaction $V$ between the immobile
particles. Thus the total Hamiltonian of the system reads:

\begin{eqnarray}
H_{0}  =  H_{FK} + V,
\label{H0} \\
H_{FK}  =  -t \sum\limits_{\langle x,y \rangle_{1}} \, \left(
c^{+}_{x}c_{y}+c^{+}_{y}c_{x} \right) + U\sum\limits_{x}\left(
c^{+}_{x}c_{x} - \frac{1}{2} \right) s_{x} ,
\label{HFK} \\
V=\frac{W}{8} \sum\limits_{\langle x,y \rangle_{1}} s_{x}s_{y}
-\frac{\tilde{\varepsilon}}{16} \sum\limits_{\langle x,y
\rangle_{2}} s_{x}s_{y}. \label{V}
\end{eqnarray}

In the above formulae, the underlying lattice is a square lattice,
denoted $\Lambda$, consisting of sites $x, y, \ldots$, whose
number is $|\Lambda|$, having the shape of a $\sqrt{|\Lambda|}
\times \sqrt{|\Lambda|}$ torus. In (\ref{HFK},\ref{V}) and below,
the sums $\sum_{\langle x,y \rangle_{i}}$, $i=1,2,3$, stand for
the summation over all the $i$-th order \nn{} pairs of lattice
sites in $\Lambda$, with each pair counted once.

The subsystem of mobile spinless particles is described in terms of
creation and annihilation operators of an electron at site $x$:
$c^{+}_{x}$, $c_{x}$, respectively, satisfying the canonical
anticommutation relations (spinless electrons) or the commutation
relations of spin $1/2$ operators $S^{+}_{x}, S^{-}_{x}, S^{z}_{x}$
(hardcore bosons). The total electron-number (boson-number) operator
is $N_{e} = \sum_{x} c^{+}_{x} c_{x}$, and (with a little abuse of
notation) the corresponding electron (boson) density is $\rho_{e} =
N_{e} /|\Lambda|$. There is no direct interaction between mobile
particles. Their energy is due to hopping, with $t$ being the \nn{}
hopping intensity, and due to the interaction with the localized
particles, whose strength is controlled by the coupling constant
$U$.

The subsystem of localized particles (here-after called ions), is
described by a collection of pseudo-spins $\left\{ s_{x} \right\}_{x
\in \Lambda}$, with $s_{x} = 1, -1$ ($s_{x}=1$ if the site $x$ is
occupied by an ion and $s_{x}=-1$ if it is empty), called the {\em
ion configurations}. The total number of ions is $N_{i} = \sum_{x} (
s_{x} + 1 )/2$ and the ion density is $\rho_{i} = N_{i}/|\Lambda|$.
In our model the ions interact directly, with energy given by $V$.

Clearly, in the composite system, whose Hamiltonian is given by
(\ref{H0}), with arbitrary electron-ion (boson-ion) coupling $U$,
the particle-number operators $N_{e}$, $N_{i}$, and pseudo-spins
$s_{x}$, are conserved. Therefore the description of the classical
subsystem in terms of the ion configurations $S =\left\{ s_{x}
\right\}_{x \in \Lambda}$ remains valid. Whenever periodic
configurations of pseudo-spins are considered, it is assumed that
$\Lambda$ is sufficiently large, so that it accommodates an integer
number of elementary cells.

Nowadays, $H_{FK}$ is widely known as the Hamiltonian of the
spinless Falicov--Kimball model, a simplified version of the
Hamiltonian put forward in \cite{FK}. To the best of our knowledge,
the spinless-fermion Falicov--Kimball model is the unique system of
interacting fermions, for which the existence of a long-range order
(the chessboard phase) \cite{brandt,kennedy1} and of a phase
separation (the segregated phase) \cite{lieb1,lieb2} have been
proved. A review of other rigorous results and an extensive list of
relevant references can be found in \cite{GM,JL}).

In what follows, we shall study the ground-state phase diagram of
the system defined by (\ref{H0}) in the grand-canonical ensemble.
That is, let
\begin{equation}
H \left( \mu_{e}, \mu_{i} \right) = H_{0} - \mu_{e}N_{e} -
\mu_{i}N_{i}, \label{Hmu}
\end{equation}
where $\mu_{e}$, $\mu_{i}$ are the chemical potentials of the
electrons (bosons) and ions, respectively, and let $E_{S}\left(
\mu_{e}, \mu_{i} \right)$ be the ground-state energy of $H \left(
\mu_{e}, \mu_{i} \right)$, for a given configuration $S$ of the
ions. Then, the ground-state energy of $H \left( \mu_{e}, \mu_{i}
\right)$, $E_{G}\left( \mu_{e}, \mu_{i} \right)$,  is defined as
$E_{G}\left( \mu_{e}, \mu_{i} \right) = \min \left\{ E_{S} \left(
\mu_{e}, \mu_{i} \right): S \right\}$. The minimum is attained at
the set $G$ of the ground-state configurations of ions. We shall
determine the subsets of the $\left( \mu_{e}, \mu_{i}
\right)$-plane, where $G$ consists of periodic configurations of
ions, uniformly in the size of the underlying square lattice.

In studies of grand-canonical phase diagrams an important role is
played by unitary transformations ({\em hole--particle
transformations}) that exchange particles and holes,
$c^{+}_{x}c_{x} \rightarrow 1 - c^{+}_{x}c_{x}$, $s_{x}
\rightarrow -s_{x}$, and for some $\left( \mu^{0}_{e}, \mu^{0}_{i}
\right)$ leave the Hamiltonian  $H \left( \mu_{e}, \mu_{i}
\right)$ invariant. For the mobile particles such a role is played
by the transformations: $c_{x}^{+} \rightarrow \epsilon_{x}
c_{x}$, where $\epsilon_{x} = 1$ for bosons while for fermions
$\epsilon_{x} = 1$ at the even sublattice of $\Lambda$ and
$\epsilon_{x} = -1$ at the odd one. Clearly, since $H_{0}$ is
invariant under the joint hole--particle transformation of mobile
and localized particles, $H \left( \mu_{e}, \mu_{i} \right)$ is
hole--particle invariant at the point $(0,0)$. At the
hole--particle symmetry point, the system under consideration has
very special properties, which simplify studies of its phase
diagram \cite{kennedy1}. Moreover, by means of the defined above
hole-particle transformations one can determine a number of
symmetries of the grand-canonical phase diagram \cite{GJL}. The
peculiarity of the model is that the case of attraction ($U<0$)
and the case of repulsion ($U>0$) are related by a unitary
transformation (the hole-particle transformation for ions): if $S
=\left\{ s_{x} \right\}_{x \in \Lambda}$ is a ground-state
configuration at $\left( \mu_{e}, \mu_{i} \right)$ for $U>0$, then
$-S =\left\{- s_{x} \right\}_{x \in \Lambda}$ is the ground-state
configuration at $\left( \mu_{e}, -\mu_{i} \right)$ for $U<0$.
Consequently, without any loss of generality one can fix the sign
of the coupling constant $U$. Moreover (with the sign of $U$
fixed), there is an {\em inversion symmetry\/} of the
grand-canonical phase diagram, that is, if $S$ is a ground-state
configuration at $\left( \mu_{e}, \mu_{i} \right)$, then $-S$ is
the ground-state configuration at $\left( -\mu_{e}, -\mu_{i}
\right)$. Therefore, it is enough to determine the phase diagram
in a half-plane specified by fixing the sign of one of the
chemical potentials.

Our aim in this paper is to investigate the ground-state phase
diagrams of our systems, for general values of the energy
parameters that appear in $H \left( \mu_{e}, \mu_{i} \right)$.
According to the state of art, this is feasible only in the
strong-coupling regime, \ie{} when $|t/U|$ is sufficiently small.
Therefore, from now on we shall consider exclusively the case of a
large positive coupling $U$, and we express all the parameters of
$H \left( \mu_{e}, \mu_{i} \right)$ in the units of $U$,
preserving the previous notation.

In the strong-coupling regime, the ground-state energy $E_{S}
\left( \mu_{e}, \mu_{i} \right)$ can be expanded in a power series
in $t$. One of the ways to achieve this, for fermions and for
hardcore bosons, is a closed-loop expansion \cite{MM-S,GMMU}. The
result, with the expansion terms up to order four shown
explicitly, reads:

\begin{eqnarray}
\label{Expfmn} E^{fermion}_{S}\left( \mu_{e}, \mu_{i} \right) &=&
-\frac{1}{2} \left( \mu_{i}-\mu_{e} \right) \sum\limits_{x} s_{x}
-\frac{1}{2} \left( \mu_{i}+\mu_{e}+1 \right) |\Lambda| +
\nonumber \\
&&\left[ \frac{t^{2}}{4}- \frac{9t^{4}}{16} + \frac{W}{8} \right]
\sum\limits_{\langle x,y \rangle_{1}}  s_{x}s_{y} + \left[
\frac{3t^{4}}{16} -\frac{\tilde{\varepsilon}}{16} \right]
\sum\limits_{\langle x,y \rangle_{2}} s_{x}s_{y} +
\nonumber \\
&&\frac{t^{4}}{8} \sum\limits_{\langle x,y \rangle_{3}} s_{x}s_{y} +
\frac{t^{4}}{16} \sum\limits_{P} \left(1+5s_{P}\right) + R^{(4)},
\end{eqnarray}
for fermions, and
\begin{eqnarray}
\label{Exphcb} E^{boson}_{S}\left( \mu_{e}, \mu_{i} \right) &=&
-\frac{1}{2} \left( \mu_{i}-\mu_{e} \right) \sum\limits_{x} s_{x}
-\frac{1}{2} \left( \mu_{i}+\mu_{e}+1 \right) |\Lambda| +
\nonumber \\
&&\left[ \frac{t^{2}}{4}- \frac{5t^{4}}{16} + \frac{W}{8} \right]
\sum\limits_{\langle x,y \rangle_{1}}  s_{x}s_{y} + \left[
\frac{5t^{4}}{16} -\frac{\tilde{\varepsilon}}{16} \right]
\sum\limits_{\langle x,y \rangle_{2}} s_{x}s_{y} +
\nonumber \\
&&\frac{t^{4}}{8} \sum\limits_{\langle x,y \rangle_{3}} s_{x}s_{y} -
\frac{t^{4}}{16} \sum\limits_{P} \left(5+s_{P}\right) +
{\tilde{R}}^{(4)},
\end{eqnarray}
for hardcore bosons, up to a term independent of the ion
configuration and the chemical potentials. In (\ref{Expfmn}) and
(\ref{Exphcb}), $P$ denotes the $(2 \times 2)$-plaquette of the
square lattice $\Lambda$, $s_{P}$ stands for the product of
pseudo-spins assigned to the corners of $P$, and the remainders
$R^{(4)}$, ${\tilde{R}}^{(4)}$, which are independent of the
chemical potentials and the parameters $W$ and
$\tilde{\varepsilon}$, collect those terms of the expansion that
are proportional to $t^{2m}$, with $m=3,4,\ldots$. It can be
proved that the above expansions  are absolutely convergent,
uniformly in $\Lambda$, provided that $t < 1/16$ and $\left|
\mu_{e} \right| < 1 - 16t$ \cite{MM-S,GMMU}. Moreover, under these
conditions the particle densities satisfy the half-filling
relation: $\rho_{e} + \rho_{i} =1$.

\section{Phase diagrams according to truncated expansions}

We note that on taking into account the inversion symmetry of the
phase diagram in the $\left( \mu_{e}, \mu_{i} \right)$-plane and
the fact that the ground-state energies depend only on the
difference of the chemical potentials, in order to determine the
phase diagram in the stripe $\left| \mu_{e} \right| < 1 - 16t$ it
is enough to consider the phase diagram at the half-line
$\mu_{e}=0$, $\mu_{i} < 0$ (or $\mu_{i} >0$). At this half-line we
set $\mu_{i} \equiv \mu$.

Due to the convergence of the expansions (\ref{Expfmn}) and
(\ref{Exphcb}), it is possible to establish rigorously a part of the
phase diagram (that is the ground-state configurations of ions are
determined everywhere in the $\left( \mu_{e}, \mu_{i}
\right)$-plane, except some small regions), by determining the phase
diagram of the expansion truncated at the order $k$, that is
according to the $k$-th order effective Hamiltonians
$(E^{fermion}_{S})^{(k)} \left( 0, \mu \right)$ and
$(E^{boson}_{S})^{(k)} \left( 0, \mu \right)$. In order to construct
a phase diagram according to a  $k$-th order effective Hamiltonian,
we use the $m$-potential method introduced in \cite{Slawny}, with
technical developments described in \cite{GJL,kennedy2,GMMU}.

\subsection{The zeroth and second order ground-state phase diagrams}

An inspection of (\ref{Expfmn}) and (\ref{Exphcb}) reveals that up
to the second order the effective Hamiltonians for fermions and
hardcore bosons are the same. Hence, the discussion that follows
applies to the both cases, and the common effective Hamiltonians are
denoted as $E_{S}^{(0)} \left( 0, \mu \right)$, $E_{S}^{(2)} \left(
0, \mu \right)$.

\begin{figure}
\includegraphics[width=0.47\textwidth]{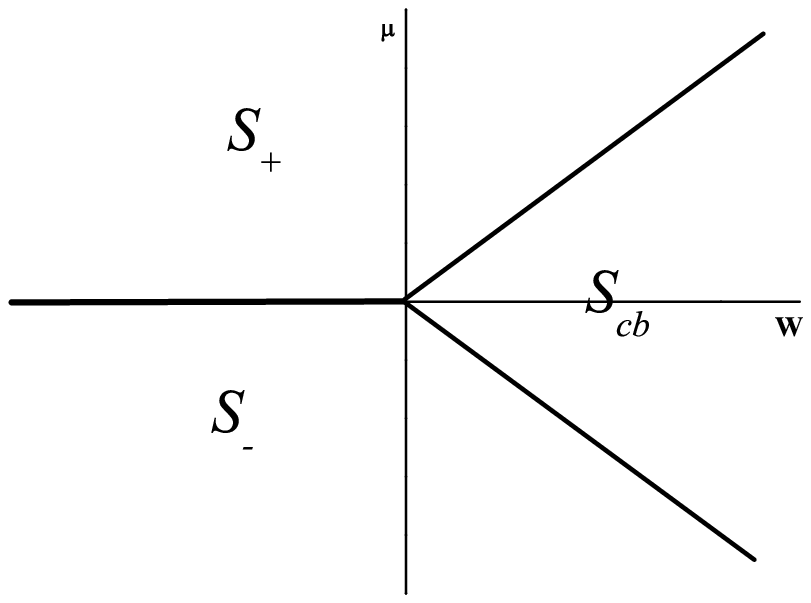}
\hfill
\includegraphics[width=0.47\textwidth]{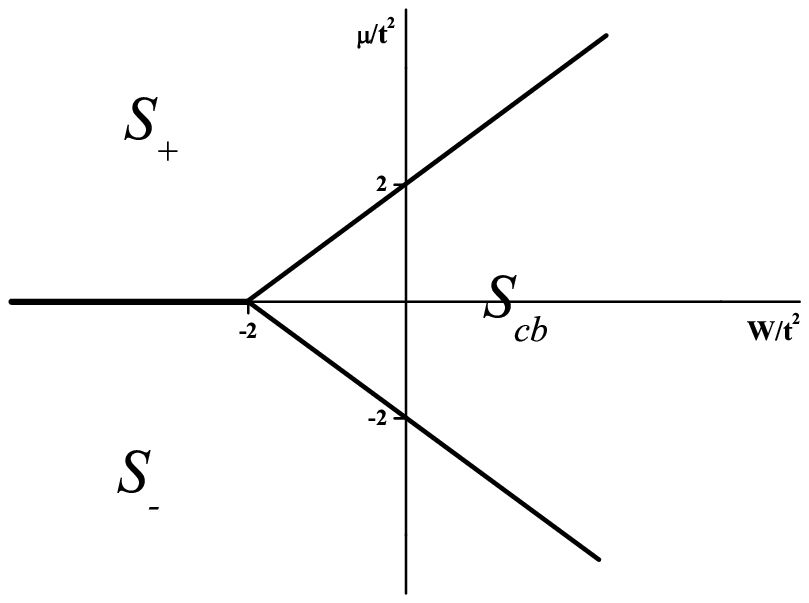}
\\
\parbox[t]{0.47\textwidth}{\caption{{\small{Ground-state
phase diagram of $E^{(0)}_{S} \left( 0, \mu \right)$ for fermion and
hard-core boson systems.}}} \label{t0}} \hfill
\parbox[t]{0.47\textwidth}{\caption{{\small{Ground-state
phase diagram of  $E^{(2)}_{S} \left( 0, \mu \right)$ for fermion
and hard-core boson systems.}}} \label{t2}}
\end{figure}

In the zeroth order,
\begin{eqnarray}
\label{E0} E_{S}^{(0)} \left( 0, \mu \right) & = & -\frac{\mu}{2}
\sum\limits_{x} \left( s_{x} +1 \right) + \frac{W}{8}
\sum\limits_{\langle x,y \rangle_{1}} s_{x}s_{y}
-\frac{\tilde{\varepsilon}}{16}
\sum\limits_{\langle x,y \rangle_{2}} s_{x}s_{y} \nonumber \\
&=& \sum\limits_{P} H_{P}^{(0)},
\end{eqnarray}
where
\begin{eqnarray}
\label{HP0} H_{P}^{(0)} =  -\frac{\mu}{8}
\sideset{}{'}\sum\limits_{x} \left( s_{x} +1 \right) + \frac{W}{16}
\sideset{}{'} \sum\limits_{\langle x,y \rangle_{1}} s_{x}s_{y}
-\frac{\tilde{\varepsilon}}{16} \sideset{}{'} \sum\limits_{\langle
x,y \rangle_{2}} s_{x}s_{y},
\end{eqnarray}
and the primed sums in (\ref{HP0}) are restricted to a plaquette
$P$.  For any $\tilde{\varepsilon} > 0$, the plaquette potentials
$H_{P}^{(0)}$ are minimized by the restrictions to a plaquette $P$
of a few periodic configurations of ions on $\Lambda$. Namely,
$S_{-}$ --- the empty configuration ($S_{+}$ --- the completely
filled configuration), where $s_{x}=-1$ at every site ($s_{x}=+1$
at every site), and the chessboard configurations $S_{cb}^{e}$,
where $s_{x}=\epsilon_{x}$, and $S_{cb}^{o}$ (where
$s_{x}=-\epsilon_{x}$), with $\epsilon_{x}=1$ if $x$ belongs to
the even sublattice of $\Lambda$ and $\epsilon_{x}=-1$ otherwise.
Moreover, out of the restrictions of those configurations to a
plaquette $P$, $S_{-|P}$, $S_{+|P}$, $S^{e}_{cb|P}$, and
$S_{cb|P}^{o}$, only four ground-state configurations on $\Lambda$
can be built, which coincide with the four configurations named
above. Clearly, this is due to the direct \nnn{} attractive
interaction between the ions. The section of the phase diagram in
the $(W,\mu,\tilde{\varepsilon})$ space by a plane with
$\tilde{\varepsilon} > 0$, according to the effective Hamiltonian
$E^{(0)}_{S} \left( 0, \mu \right)$, is shown in Fig.~\ref{t0}.

In the second order,
\begin{eqnarray}
\label{E2} E_{S}^{(2)} \left( 0, \mu \right) & = & -\frac{\mu}{2}
\sum\limits_{x} \left( s_{x} +1 \right) + \left[ \frac{t^{2}}{4}+
\frac{W}{8} \right]\sum\limits_{\langle x,y \rangle_{1}} s_{x}s_{y}
-\frac{\tilde{\varepsilon}}{16} \sum\limits_{\langle x,y
\rangle_{2}} s_{x}s_{y}
\nonumber \\
&=& \sum\limits_{P} H_{P}^{(2)},
\end{eqnarray}
where $H_{P}^{(2)}$,
\begin{eqnarray}
\label{H2} H_{P}^{(2)} =  -\frac{\mu}{8} \sideset{}{'}
\sum\limits_{x} \left( s_{x} +1 \right) + \left[\frac{t^{2}}{8}+
\frac{W}{16} \right] \sideset{}{'} \sum\limits_{\langle x,y
\rangle_{1}} s_{x}s_{y} -\frac{\tilde{\varepsilon}}{16}
\sideset{}{'} \sum\limits_{\langle x,y \rangle_{2}} s_{x}s_{y},
\end{eqnarray}
and the primed sums are restricted to a plaquette $P$. Since
$E_{S}^{(2)}$ differs from $E_{S}^{(0)}$ by the strength of \nn{}
interactions, the  only effect on the phase diagram is that the
coexistence lines are translated by the vector $(-2t^{2},0,0)$.
The section of the phase diagram in the
$(W,\mu,\tilde{\varepsilon})$ space by a plane with
$\tilde{\varepsilon} > 0$, according to the effective Hamiltonian
$E^{(2)}_{S} \left( 0, \mu \right)$, is shown in Fig.~\ref{t2}.

\subsection{The fourth order ground-state phase diagrams}

It follows from the investigation of  the phase diagram up to the
second order that the degeneracy is finite, independently of the
size of $\Lambda$, everywhere except the coexistence lines in the
($\tilde{\varepsilon}=0$)-plane, where it grows exponentially with
$|\Lambda|$. Only there the effect of the fourth order
interactions can be most significant. The meeting point of the
coexistence lines in the ($\tilde{\varepsilon}=0$)-plane, where
the coexistence line of $S_{+}$ and $S_{-}$ sticks to the
stability domain of chessboard configurations, appears to be
particularly interesting. In that point, the energies of all the
configurations are the same. In what follows, we shall study the
phase diagrams of spinless fermions and spinless hardcore bosons
up to the fourth order, in a neighborhood of radius $O(t^{4})$ of
the point $(-2t^{2},0,0)$. In this neighborhood, it is convenient
to introduce new coordinates, $(W,\mu,\tilde{\varepsilon}) \to
(\omega ,\delta,\varepsilon)$,
\begin{eqnarray}
\label{new-var} W = -2t^{2}+ t^{4}\omega , \hspace{5mm}
\mu = t^{4} \delta , \hspace{5mm}
\tilde{\varepsilon} = t^{4} \varepsilon ,
\end{eqnarray}
and a new (equivalent to $(E^{fermion}_{S})^{(4)}$) $t$-independent
effective Hamiltonian, $(H^{fermion}_{{\mathrm{eff}}})^{(4)}$,
\begin{eqnarray}
\label{Heff-1} (E^{fermion}_{S})^{(4)} \left( 0, \delta \right) =
\frac{t^{4}}{2} (H^{fermion}_{{\mathrm{eff}}})^{(4)}.
\end{eqnarray}
Then, in the spirit of the $m$-potential method, we express
$(H^{fermion}_{{\mathrm{eff}}})^{(4)}$ by the potentials
$(H^{fermion}_{T})^{(4)}$,
\begin{equation}
\label{HT} (H^{fermion}_{{\mathrm{eff}}})^{(4)} = \sum\limits_{T}
(H^{fermion}_{T})^{(4)},
\end{equation}
where, in  terms of the new variables,
\begin{eqnarray}
\label{H4fmn-1} (H^{fermion}_{T})^{(4)} &=& -\delta \left( s_{5} +1
\right) + \frac{1}{24} \left( \omega-\frac{9}{2} \right)
\sideset{}{''}\sum\limits_{\langle x,y \rangle_{1}} s_{x}s_{y} +
\frac{1}{32} \left(3 - \varepsilon \right)
\sideset{}{''} \sum\limits_{\langle x,y \rangle_{2}} s_{x}s_{y} + \nonumber \\
& & \frac{1}{12} \sideset{}{''} \sum\limits_{\langle x,y
\rangle_{3}} s_{x}s_{y} + \frac{1}{32} \sideset{}{''}
\sum\limits_{P} \left( 5s_{P}+1 \right) .
\end{eqnarray}
In (\ref{HT}) and (\ref{H4fmn-1}), $T$ stands for a $(3 \times
3)$-plaquette (later on called the $T$-plaquette) of a square
lattice, whose sites are labeled from the left to the right,
starting at the bottom left corner  and ending in the upper right
one. Consequently, $s_{5}$ is the pseudo-spin of the central site.
The double-primed sums are restricted to a $T$-plaquette. For
bosons, we introduce $(H^{boson}_{{\mathrm{eff}}})^{(4)}$ and
$(H^{boson}_{T})^{(4)}$ in the same manner as for fermions, with
\begin{eqnarray}
\label{H4hcb-1} (H^{boson}_{T})^{(4)} &=& -\delta \left( s_{5} +1
\right) + \frac{1}{24} \left( \omega -\frac{5}{2} \right)
\sideset{}{''} \sum\limits_{\langle x,y \rangle_{1}} s_{x}s_{y} +
\frac{1}{32} \left( 5 - \varepsilon \right)
\sideset{}{''} \sum\limits_{\langle x,y \rangle_{2}} s_{x}s_{y} + \nonumber \\
& & \frac{1}{12} \sideset{}{''} \sum\limits_{\langle x,y
\rangle_{3}} s_{x}s_{y} - \frac{1}{32} \sideset{}{''}
\sum\limits_{P} \left( s_{P}+5 \right).
\end{eqnarray}

\begin{figure}
\centering \includegraphics[width=0.9\textwidth]{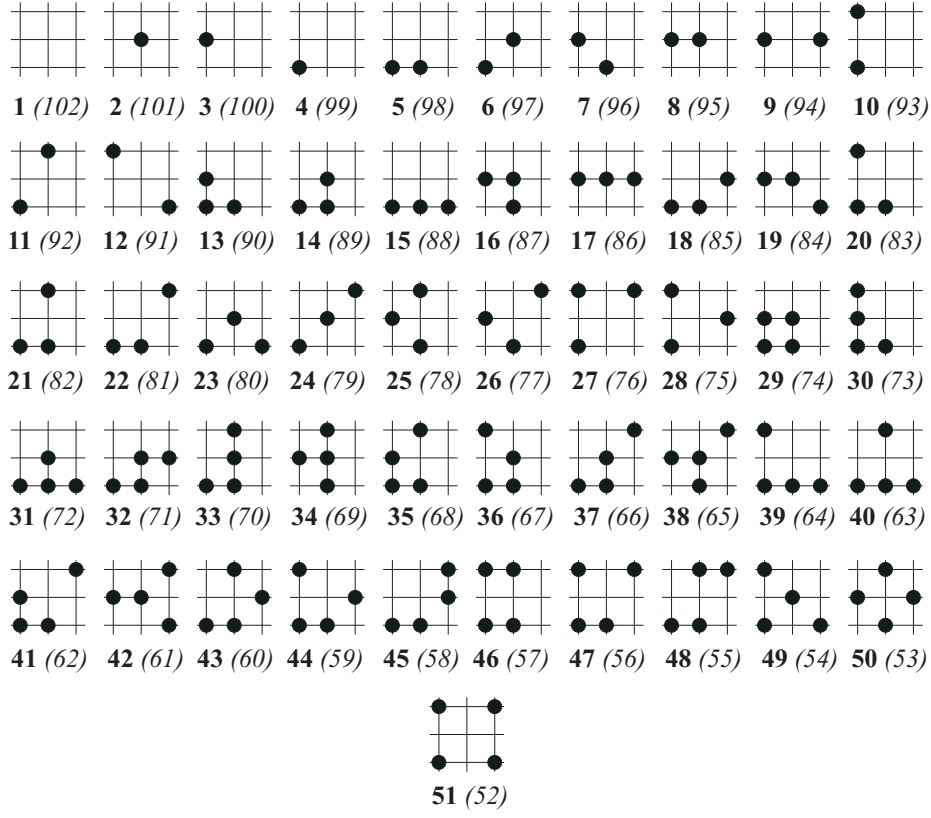}
\caption{{\small{All the $T$-plaquette configurations, up to the
symmetries of $H_{0}$. The configurations that can be obtained from
the displayed ones by the hole-particle transformation are not
shown, only the numbers assigned to them are given in brackets.}}}
\label{bk}
\end{figure}

By the remark at the begining  of this section, we have to search
for the lowest-energy configurations of $(E^{fermion}_{S})^{(4)}$
and $(E^{boson}_{S})^{(4)}$ among all the configurations.
Consequently, the potentials $(H^{fermion}_{T})^{(4)}$ and
$(H^{boson}_{T})^{(4)}$ have to be minimized over all the
$T$-plaquette configurations. There are (up to the symmetries of
$H_{0}$) 102 different $T$-plaquette configurations, shown in
Fig.~\ref{bk}.
Unfortunately, in contrast to the lower-order cases, the
potentials $(H^{fermion}_{T})^{(4)}$ and $(H^{boson}_{T})^{(4)}$
turn out to be the $m$-potentials only in a small part of
$(\omega,\delta,\varepsilon)$-space. This difficulty can be
overcome by introducing the so called zero-potentials
\cite{GJL,kennedy2,GMMU}, denoted $K_{T}^{(4)}$, that are
invariant with respect to the symmetries of $H_{0}$ and satisfy
the condition
\begin{eqnarray}
\sum\limits_{T} K_{T}^{(4)} =0. \label{KTcond}
\end{eqnarray}
Following \cite{kennedy2}, $K_{T}^{(4)}$ are chosen in the form
\begin{eqnarray}
K^{(4)}_{T}=\sum\limits_{i=1}^{5} \alpha_{i}k_{T}^{(i)},
\end{eqnarray}
where the coefficients $\alpha_{i}$, depending on
$(\omega,\delta,\varepsilon)$ in general, have to be determined in
the process of constructing the phase diagram, and the potentials
$k_{T}^{(i)}$ are defined by
\begin{eqnarray}
k_{T}^{(1)} & = & s_{1}+s_{3}+s_{7}+s_{9}-4s_{5}, \nonumber \\
k_{T}^{(2)} & = & s_{2}+s_{4}+s_{6}+s_{8}-4s_{5}, \nonumber \\
k_{T}^{(3)} & = & s_{1}s_{2}+s_{2}s_{3}+s_{3}s_{6}
+s_{6}s_{9}+s_{8}s_{9}+s_{7}s_{8}+s_{4}s_{7}+s_{1}s_{4} \nonumber\\
&& -2s_{2}s_{5}-2s_{5}s_{6}-2s_{5}s_{8}-2s_{4}s_{5}, \nonumber \\
k_{T}^{(4)} & = & s_{1}s_{5}+s_{3}s_{5}+s_{5}s_{9}
+s_{5}s_{7}-s_{2}s_{4}-s_{4}s_{8}-s_{8}s_{6}-s_{2}s_{6}, \nonumber \\
k_{T}^{(5)} & = & s_{1}s_{3}+s_{3}s_{9}+s_{7}s_{9}+s_{1}s_{7}
-2s_{4}s_{6}-2s_{2}s_{8}.
\end{eqnarray}
Note that the potentials $k_{T}^{(i)}$ are invariant with respect to
the symmetries of $H_{0}$ and they satisfy (\ref{KTcond}), therefore
these properties are shared by the potentials $K^{(4)}_{T}$.

By means of the zero-potentials $K^{(4)}_{T}$, new candidates for
$m$-potentials can be introduced,
\begin{eqnarray}
(H^{fermion}_{{\mathrm{eff}}})^{(4)} = \sum\limits_{T} \left(
(H^{fermion}_{T})^{(4)} + K^{(4)}_{T} \right),
\end{eqnarray}
with an analogous representation of
$(H^{boson}_{{\mathrm{eff}}})^{(4)}$, and now the task is to
minimize the potentials $(H^{fermion}_{T})^{(4)} + K^{(4)}_{T}$and
$(H^{boson}_{T})^{(4)} + K^{(4)}_{T}$ over all the $T$-plaquette
configurations.

In our study of the ground-state phase diagrams we limit ourselves
to the ($\delta=0$)-plane, where the both considered systems are
hole-particle invariant, and to the ($\varepsilon=0$)-plane. Our
analysis of the minima of the $T$-plaquette potentials,
$(H^{fermion}_{T})^{(4)}$ and $(H^{boson}_{T})^{(4)}$ augmented by
the zero-potentials $K^{(4)}_{T}$, shows that these planes are
partitioned into a finite number of open domains
${\mathcal{S}}_{D}$, each with its unique set of periodic
ground-state configurations on $\Lambda$, denoted also by
${\mathcal{S}}_{D}$. There is a finite number, independent of
$|\Lambda|$, of configurations in ${\mathcal{S}}_{D}$ and they are
related by the symmetries of $H_{0}$. The last two statements do
not apply to only one of the domains, denoted
${\mathcal{S}}_{d2}$, which will be described in the sequel.

The domains ${\mathcal{S}}_{D}$ are characterized as follows: at
each point $p$ ($p=(\omega,\varepsilon)$ or $p=(\omega,\delta)$)
of a domain ${\mathcal{S}}_{D}$, there exist a set of coefficients
$\left\{ \alpha_{i}(p) \right\}$ such that the corresponding
potentials are minimized by a set ${\mathcal{S}}_{TD}(p)$ of
$T$-plaquette configurations. Moreover, from the configurations in
${\mathcal{S}}_{TD}(p)$ one can construct only the configurations
in ${\mathcal{S}}_{D}$. The set of the restrictions to
$T$-plaquettes of the configurations from ${\mathcal{S}}_{D}$,
${\mathcal{S}}_{D|T}$, is contained in each set
${\mathcal{S}}_{TD}(p)$ with $p \in {\mathcal{S}}_{D}$. In
Table~\ref{tb4} of Appendix we mark by asterisk the cases, where
the set ${\mathcal{S}}_{TD}(p)$ contains, besides
${\mathcal{S}}_{D|T}$, some additional $T$-plaquette
configurations.

\begin{figure}
\centering\includegraphics[height=0.34\textheight]{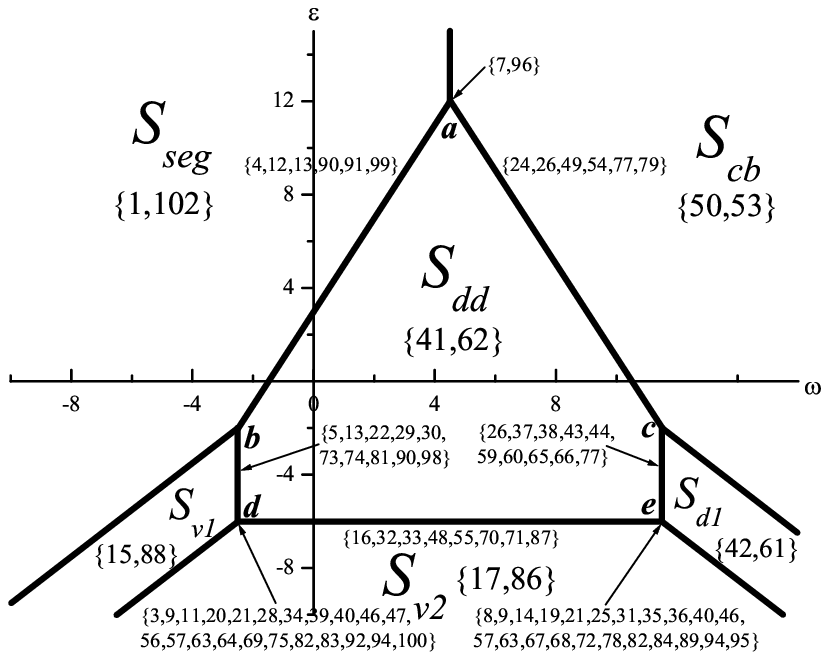}
\parbox[t]{\textwidth}{\caption{{\small{The ($\mu=0$)-phase diagram
of the effective Hamiltonian $(H^{fermion}_{{\mathrm{eff}}})^{(4)}$.
The numbers in curly brackets, displayed by the symbols of open
domains ${\mathcal{S}}_{D}$, denote the $T$-plaquette configurations
that minimize the $m$-potential in ${\mathcal{S}}_{D}$. The numbers
in curly brackets, displayed by the boundary-line segments or by the
arrows pointing towards boundary segments (or their crossing points)
identify the additional minimizing $T$-plaquette configurations. For
more comments see the text. The boundary-line segments can be
determined by means of their intersection points: ${\bf a} =
(9/2,12)$, ${\bf b} = (-5/2,-2)$, ${\bf c} = (23/2,-2)$, ${\bf d} =
(-5/2,-6)$, ${\bf e} = (23/2,-6)$, and by the slope $1$ of the
boundary of ${\mathcal{S}}_{v1}$, and the slope $-1$ of the boundary
of ${\mathcal{S}}_{d1}$.}}} \label{t40f}} \vfill
\centering\includegraphics[height=0.34\textheight]{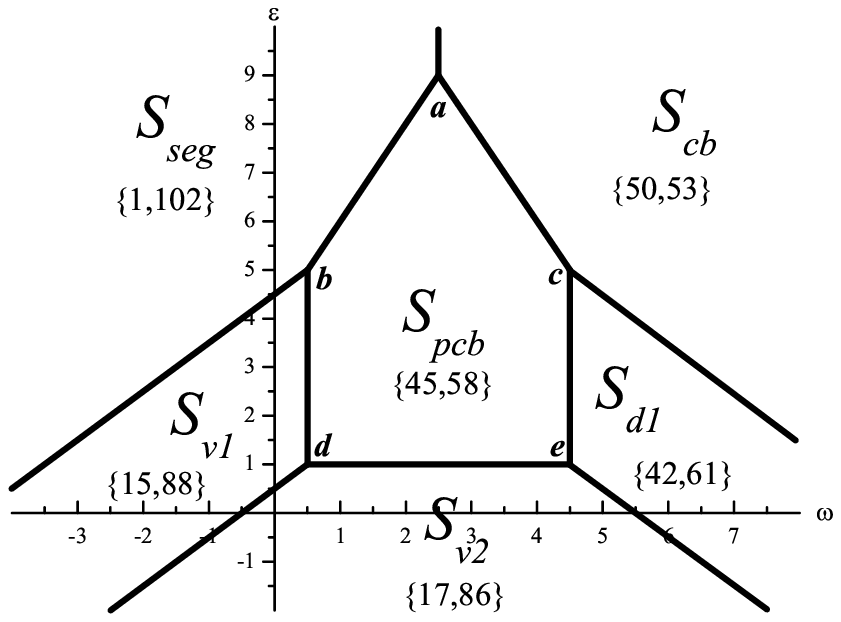}
\parbox[t]{\textwidth}{\caption{{\small{The ($\mu=0$)-phase diagram
of the effective Hamiltonian $(H^{boson}_{{\mathrm{eff}}})^{(4)}$.
The boundary-line segments can be determined by means of their
intersection points: ${\bf a} = (5/2,9)$, ${\bf b} = (1/2,5)$, ${\bf
c} = (9/2,5)$, ${\bf d} = (1/2,1)$, ${\bf e} = (9/2,1)$, and by the
slope $1$ of the boundary of ${\mathcal{S}}_{v1}$, and the slope
$-1$ of the boundary of ${\mathcal{S}}_{d1}$. For more explanations
see the description of Fig.~\ref{t40f}. }}} \label{t40b}}
\end{figure}

\begin{figure}
\centering\includegraphics[height=0.34\textheight]{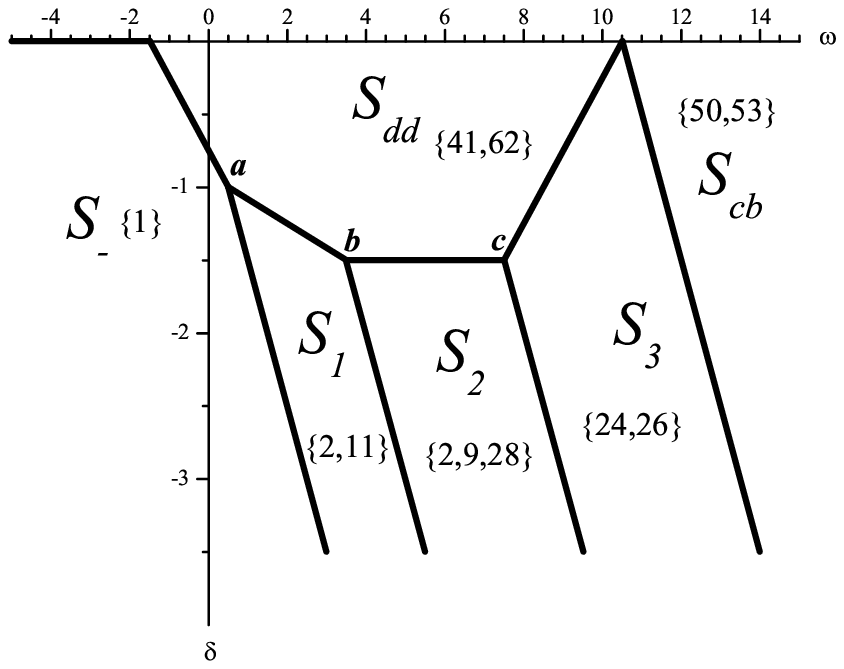}
\parbox[t]{\textwidth}{\caption{{\small{The ($\varepsilon=0$)-phase diagram
of the effective Hamiltonian $(H^{fermion}_{{\mathrm{eff}}})^{(4)}$.
The boundary-line segments can be determined by means of their
intersection points: ${\bf a} = (1/2,-1)$, ${\bf b} = (7/2,-3/2)$,
${\bf c} = (15/2,-3/2)$, and by the slope $-1$ of the boundaries of
${\mathcal{S}}_{1}$, ${\mathcal{S}}_{2}$, ${\mathcal{S}}_{3}$. For
more comments see the description of Fig.~\ref{t40f}.}}}
\label{t41f}} \vfill
\centering\includegraphics[height=0.34\textheight]{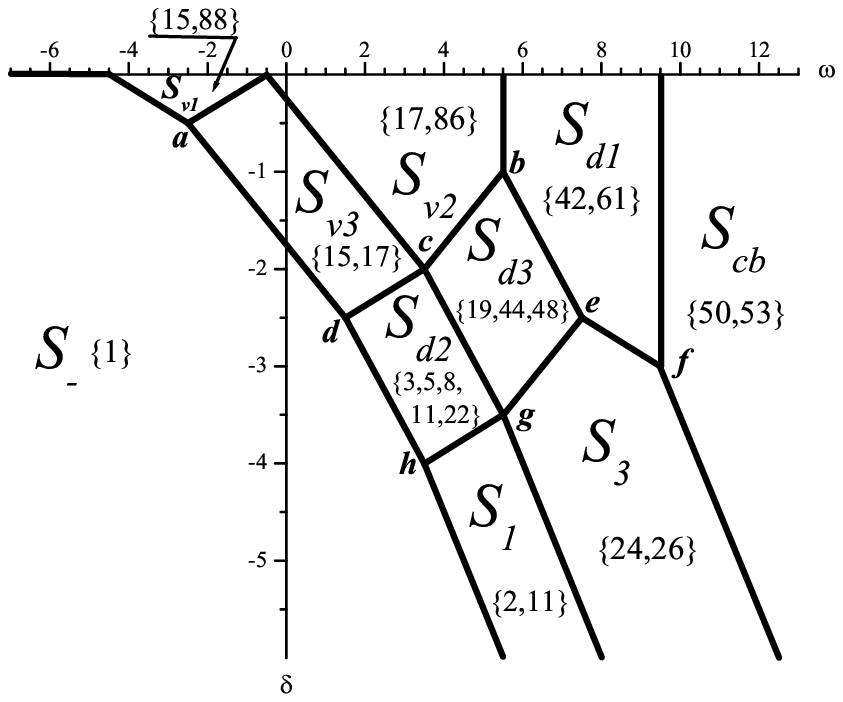}
\parbox[t]{\textwidth}{\caption{{\small{The ($\varepsilon=0$)-phase diagram
of the effective Hamiltonian $(H^{boson}_{{\mathrm{eff}}})^{(4)}$.
The boundary-line segments can be determined by means of their
intersection points: ${\bf a} = (-5/2,-1/2)$, ${\bf b} = (11/2,-1)$,
${\bf c} = (7/2,-2)$, ${\bf d} = (3/2,-5/2)$, ${\bf e} =
(15/2,-5/2)$, ${\bf f} = (19/2,-3)$, ${\bf g} = (11/2,-7/2)$, ${\bf
h} = (7/2,-4)$, and by the slope $-1$ of the boundaries of
${\mathcal{S}}_{1}$ and ${\mathcal{S}}_{3}$. For more comments see
the description of Fig.~\ref{t40f}}}} \label{t41b}}
\end{figure}

\begin{figure}
\centering \includegraphics[width=0.9\textwidth]{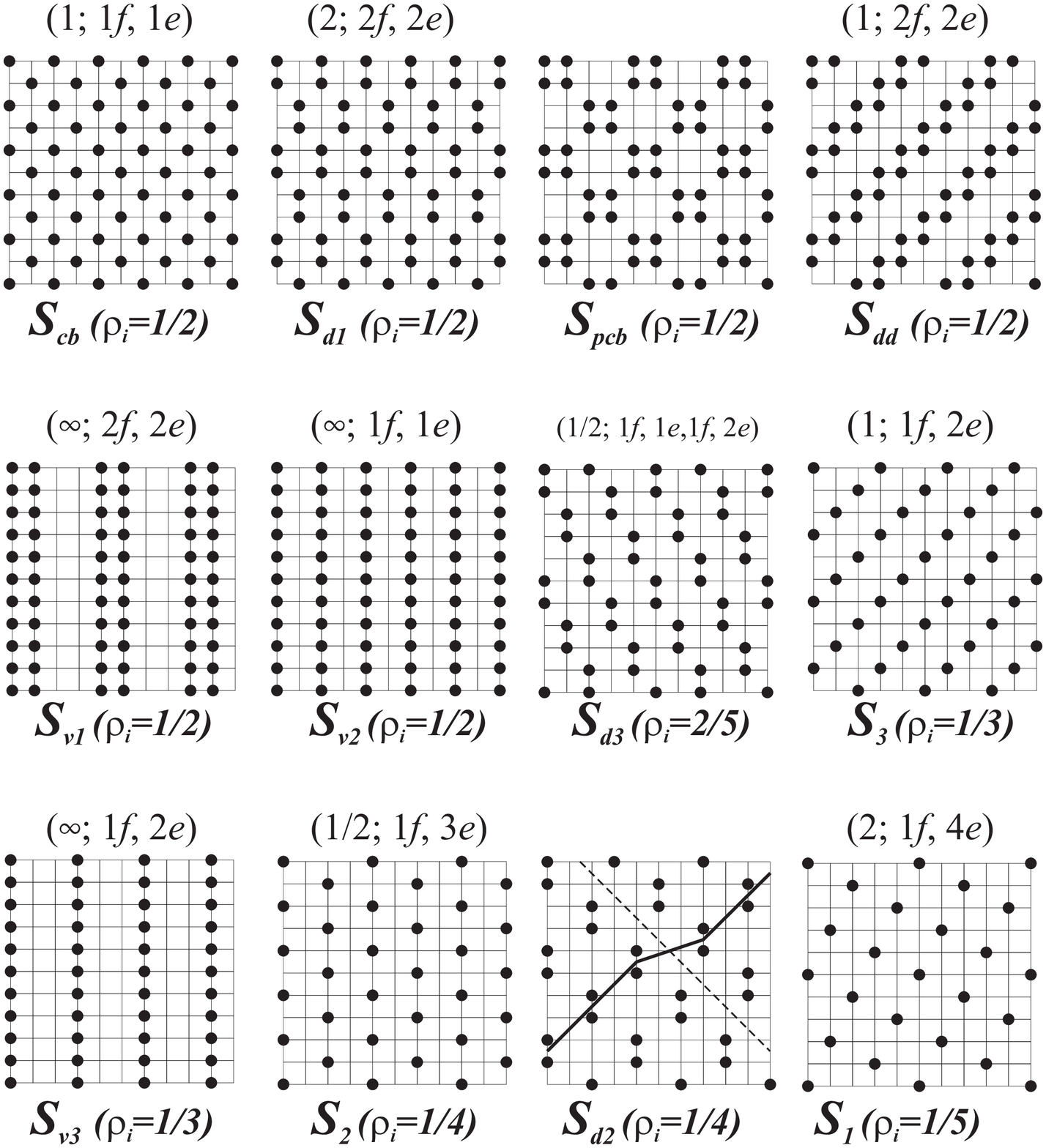}
\caption{{\small{Representative configurations of the sets
${\mathcal{S}}_{D}$  of ground-state configurations. The remaining
configurations of ${\mathcal{S}}_{D}$ can be obtained by applying
the symmetries of $H_{0}$ to the displayed configurations. As a
representative configuration of ${\mathcal{S}}_{d2}$, we show a
configuration with one defect line (the dashed line). The continuous
line is a guide for the eye. For more comments see the text.}}}
\label{conf}
\end{figure}

Specifically, for $\delta=0$  the ground-state phase diagrams due to
the effective Hamiltonians $(H^{fermion}_{{\mathrm{eff}}})^{(4)}$
and $(H^{boson}_{{\mathrm{eff}}})^{(4)}$ are shown in
Fig.~\ref{t40f} and Fig.~\ref{t40b},
respectively, while the corresponding ground-state phase diagrams for
$\varepsilon =0$, in Fig.~\ref{t41f} and Fig.~\ref{t41b}.
In Fig.~\ref{conf} we display the representatives of the sets
${\mathcal{S}}_{D}$ of ground-state configurations.
That is, the remaining configurations of
${\mathcal{S}}_{D}$ can be obtained easily by applying the
symmetries of $H_{0}$ to the displayed configurations. The domain
${\mathcal{S}}_{seg}$, having no representative in Fig.~\ref{conf},
consists of the two translation-invariant configurations $S_{+}$ and
$S_{-}$, related by the hole-particle transformation.

Only in the domain ${\mathcal{S}}_{d2}$, which appears in the phase
diagrams shown in Fig.~\ref{t41f},~\ref{t41b}, that is off the
hole-particle symmetry plain, the situation is not that simple. In
${\mathcal{S}}_{d2}$ one can distinguish two classes,
${\mathcal{S}}_{d2a}$ and ${\mathcal{S}}_{d2b}$, of periodic
configurations with parallelogram elementary cells. A configuration
in ${\mathcal{S}}_{d2a}$ consists of vertical (horizontal) dimers of
filled sites that form a square lattice, where the sides of the
elementary squares have the length $2\sqrt{2}$ and the slope $\pm
1$. In a configuration of ${\mathcal{S}}_{d2b}$, the elementary
parallelograms formed by dimers have the sides of the length
$2\sqrt{2}$ and the slope $\pm 1$, and the sides of the length
$\sqrt{10}$ and the slope $\pm 1/3$. Two configurations, one from
${\mathcal{S}}_{d2a}$ and one from ${\mathcal{S}}_{d2b}$, having the
same kind of dimers (vertical or horizontal), can be merged together
along a ``defect line'' of the slope $\pm 1$ (dashed line in
Fig.~\ref{conf}), as shown in Fig.~\ref{conf}, without increasing
the energy. By introducing more defect lines one can construct many
ground-state configurations whose number scales with the size of the
lattice as $\exp{(const \sqrt{\Lambda})}$. While for a finite
lattice all the configurations are periodic, many of them become
aperiodic in the infinite volume limit.

As mentioned above, except ${\mathcal{S}}_{d2}$ all the other sets
${\mathcal{S}}_{D}$ contain exclusively periodic configurations.
The set  ${\mathcal{S}}_{pcb}$ (of plaquette-chessboard
configurations) contains configurations built out of elementary
plaquettes with filled sites, forming a square lattice according
to the same rules as filled sites form a square lattice in the
chessboard configurations from  ${\mathcal{S}}_{cb}$. The
remaining sets  ${\mathcal{S}}_{D}$ consist of configurations that
have a quasi-one-dimensional structure. That is, they are built
out of completely filled and completely empty lattice lines of
given slope. Such a configuration can be specified by the slope of
the filled lattice lines of the representative configuration and
the succession of filled (f) and empty (e) consecutive lattice
lines in a period. For instance, the representative configuration
of ${\mathcal{S}}_{d1}$ (see Fig.~\ref{conf}) is built out of
filled lines with the slope $2$ and, in the period, two
consecutive filled lines are followed by two consecutive empty
lines, which is denoted $(2;2f,2e)$. Similar description of the
remaining quasi-one-dimensional configurations is given in
Fig.~\ref{conf}.

The coefficients $\{\alpha_{i}\}$ for which the fourth order
potentials, $(H^{fermion}_{T})^{(4)} + K^{(4)}_{T}$ and
$(H^{boson}_{T})^{(4)} + K^{(4)}_{T}$, become $m$-potentials are
given in the tables collected in the Appendix.

Finally, a remark concerning ground-state configurations at the
boundaries between the open domains ${\mathcal{S}}_{D}$ is in order.
Let ${\mathcal{S}}_{D}$ and ${\mathcal{S}}_{D^{\prime}}$ be two
domains of the considered phase diagram, sharing a boundary. At this
boundary, the set of the minimizing $T$-plaquette configurations
contains always the subset ${\mathcal{S}}_{TD} \cup
{\mathcal{S}}_{TD^{\prime}}$, but it may contain also some
additional $T$-plaquette configurations of minimal energy.
Consequently, the set of the ground-state configurations at the
boundary contains always the subset ${\mathcal{S}}_{D} \cup
{\mathcal{S}}_{D^{\prime}}$ and typically a great many of other
ground-state configurations, whose number grows indefinitely with
the size of the lattice. In the considered diagrams, the only
exception is the boundary between ${\mathcal{S}}_{seg}$ and
${\mathcal{S}}_{cb}$, where the set of the ground-state
configurations amounts exactly to ${\mathcal{S}}_{seg} \cup
{\mathcal{S}}_{cb}$.

\section{Discussion of the phase diagrams and conclusions}

We start with a few comments concerning the phase diagrams due to
the fourth order effective interactions. In the fourth order
effective interaction, the parameters $\omega$ and $\varepsilon$
control the strength of \nn{} and \nnn{} interactions, respectively.
The \nn{} interaction is repulsive and favors the chessboard
configurations if $\omega > 9/2$, for fermions ($\omega
> 5/2$, for bosons). In the opposite case (attraction) it favors
the $S_{+}$, $S_{-}$ configurations. In turn, the \nnn{} interaction
is attractive if $\varepsilon > 3$, for fermions ($\varepsilon > 5$,
for bosons), and then it reinforces the tendency towards chessboard
and uniform, $S_{+}$, $S_{-}$, configurations. When it becomes
repulsive, it frustrates the \nn{} interaction: the more negative it
is the larger $\omega$ is needed to stabilize the chessboard
configurations and the smaller $\omega$ is needed to stabilize the
$S_{+}$, $S_{-}$ configurations. Therefore, whatever the value of
$\varepsilon$ is, there is a sufficiently large $\omega$ such that
$(\varepsilon,\omega)\in {\mathcal{S}}_{cb}$ and a sufficiently
small $\omega$ such that $(\varepsilon,\omega)\in
{\mathcal{S}}_{seg}$.

Apparently, the fourth order phase diagrams in the cases of hoping
fermions and hoping bosons are quite similar if we compare the
geometry of the phase boundaries and the sets of ground states. The
main difference is in the domain occupying the central position: in
the case of fermions the ground-state configurations are
diagonal-stripe configurations, ${\mathcal{S}}_{dd}$, while in the
case of bosons this is the set ${\mathcal{S}}_{pcb}$ of
plaquette-chessboard configurations. These two sets of
configurations are ground-state configurations of the corresponding
fourth-order effective interactions with the site, \nn{}, and \nnn{}
terms dropped, which corresponds to the points $(9/2,3) \in
{\mathcal{S}}_{dd}$ and $(5/2,5) \in {\mathcal{S}}_{pcb}$. This
means that the fourth order phase diagrams can be looked upon as the
result of perturbing an Ising-like Hamiltonian, that consists of
pair interactions at distance of two lattice constants and plaquette
interactions, by \nn{} and \nnn{} interactions. More generally, it is
easy to verify that the set of ground-state configurations of the
Ising-like Hamiltonian,
\begin{eqnarray}
\label{} \sum\limits_{\langle x,y \rangle_{3}} s_{x}s_{y} + \gamma
\sum\limits_{P}  s_{P} ,
\end{eqnarray}
amount to ${\mathcal{S}}_{dd}$ if $\gamma > 0$, and to
${\mathcal{S}}_{pcb}$ if $\gamma < 0$. These ground states are
stable with respect to perturbations by \nn{} and \nnn{}
interactions, if the corresponding interaction constants are in a
certain vicinity of zero. Otherwise, new sets of ground states,
shown in Fig.~\ref{t40f} and Fig.~\ref{t40b} emerge.

The basic question to be answered, before discussing the phase
diagrams due to the complete interaction, refers to the relation
between these phase diagrams and the diagrams due to the truncated
effective interactions, obtained in the previous section.

Firstly, we note, by inspection of the phase boundaries in
Fig.~\ref{t40f} and Fig.~\ref{t40b}, that the $T$-plaquette
configurations $\{45,58\}$ are missing in the phase diagram
corresponding to fermion mobile particles. In turn, the
$T$-plaquette configurations $\{41,62\}$ are missing in the phase
diagram corresponding to boson mobile particles. Consequently, the
ground-state configurations of ${\mathcal{S}}_{pcb}$ cannot appear
not only in the fourth order fermion phase diagram but also in the
fermion phase diagram of the complete interaction. Similarly, the
ground-state configurations of ${\mathcal{S}}_{dd}$ do not appear in
boson phase diagrams.

Secondly, by adapting the arguments presented in
\cite{kennedy2,GMMU}, we can demonstrate, see for instance
\cite{DJ}, that if the remainder $R^{(4)}$ is taken into account,
then there is a (sufficiently small) constant $t_{0}$ such that for
$t<t_{0}$ the phase diagram looks the same as the phase diagram due
to the effective interaction truncated at the fourth order, except
some regions of width $O(t^{6})$, located along the boundaries between
the domains, and except the domain ${\mathcal{S}}_{d2}$. For
$t<t_{0}$ and each domain ${\mathcal{S}}_{D}$, ${\mathcal{S}}_{D}
\neq {\mathcal{S}}_{d2}$, there is a nonempty two-dimensional open
domain ${\mathcal{S}}_{D}^{\infty}$ that is contained in the
domain ${\mathcal{S}}_{D}$ and such that in
${\mathcal{S}}_{D}^{\infty}$ the set of ground-state configurations
coincides with ${\mathcal{S}}_{D}$.

Now, consider our systems for specified particle densities,
$\rho_{e} = \rho_{i} = 1/2$. According to the phase diagrams in
the hole-particle symmetry plane ($\mu =\delta =0$), shown in
Fig.~\ref{t40f} and Fig.~\ref{t40b}, if $(\varepsilon,\omega)$ is
in ${\mathcal{S}}_{seg}^{\infty}$, then the ground state is a
phase-separated state, where ion configurations are  mixtures of
$S_{+}$ and $S_{-}$ configurations, called the segregated phase.
Another phase-separated state, where the ion configurations are
mixtures of $S_{+}$, $S_{-}$, $S_{cb}^{e}$, and $S_{cb}^{o}$
configurations, exists at a line that is a small ($O(t^{2})$)
distortion of the boundary between the domains
${\mathcal{S}}_{seg}$ and ${\mathcal{S}}_{cb}$ \cite{kennedy2}. In
all the other domains ${\mathcal{S}}_{D}^{\infty}$ of these
diagrams the ground-state phase exhibits a crystalline long-range
order of ions.

By the properties of the fourth order phase diagrams, whatever the
value of $\varepsilon$ is, there is a sufficiently large $\omega$
such that the systems are in the chessboard phase
($(\varepsilon,\omega)\in {\mathcal{S}}_{cb}^{\infty}$) and a
sufficiently small $\omega$ such that the systems are in the
segregated phase ($(\varepsilon,\omega)\in
{\mathcal{S}}_{seg}^{\infty}$). In between, the systems undergo a
series of phase transitions and visit various striped phases.
Consider first unbiased systems ($\varepsilon =0$). Then, it follows
from the phase diagram in Fig.~\ref{t40f}, that the fermion system
visits necessarily the diagonal striped phase, described by the
configurations in ${\mathcal{S}}_{dd}$. In turn, the phase diagram
in Fig.~\ref{t40b} implies that the boson system has to visit the
dimerized-chessboard phase, described by the configurations in
${\mathcal{S}}_{d1}$, then the vertical/horizontal striped phase,
described by the configurations in ${\mathcal{S}}_{v2}$, and after
that a dimerized vertical/horizontal striped phase, described by the
configurations in ${\mathcal{S}}_{v1}$. These scenarios are
preserved, if $\varepsilon \in (-2 + O(t^{2}),12 - O(t^{2}))$ in the
fermion case, and $\varepsilon \in (- \infty,1 - O(t^{2}))$ in the
boson case.

Note that the same sequence of transitions, as found in the unbiased
boson system, can be realized by the fermion system if $\varepsilon$
is sufficiently small ($\varepsilon < -6 - O(t^{2})$).

For somewhat larger values of $\varepsilon$, $\varepsilon \in (-6 +
O(t^{2}),-2 - O(t^{2}))$, on its way from the chessboard phase to the
segregated phase the fermion systems visits the dimerized chessboard
phase, then the diagonal striped phase, and after that the dimerized
vertical/horizontal striped phase. The vertical/horizontal striped
phase is closer to the segregated phase than the diagonal striped
phase. In case the both kinds of stripe phases (vertical/horizontal
and diagonal) appear in a phase diagram, such a succession of
striped phases is perhaps generic, see also \cite{lemanski1,smith}.

For sufficiently large $\varepsilon$, the tendency toward the
chessboard and the uniform configurations is so strong that the both
systems jump directly from the chessboard phase to the segregated
phase.

Above, we have described the states of the considered systems for
typical values of the control parameter $\omega$. Only in small
intervals (whose width is of the order of $O(t^{2})$) of values of
$\omega$, about the transition points of the diagrams in
Fig.~\ref{t40f} and Fig.~\ref{t40b}, the states of the systems
remain undetermined.

For completeness we present also ground-state phase diagrams of the
both systems off the hole-particle symmetry plane, for
$\varepsilon=0$, see Fig.~\ref{t41f},~\ref{t41b}. Away from the
hole-particle symmetry plane new phases appear. In particular we
find the phases described by the quasi-one-dimensional
configurations ${\mathcal{S}}_{1}$, ${\mathcal{S}}_{2}$, and
${\mathcal{S}}_{3}$, well known from the studies of the phase
diagram of the spinless Falicov--Kimball model \cite{GMMU}. In the
boson phase diagram, Fig.~\ref{t41b}, there appears the domain
${\mathcal{S}}_{d2}$ that is not amenable to the kind of arguments
used in this paper.

\section{Summary}

In this paper, we address the problem of the stability of
charge-stripe phases versus the phase separated state --- the
segregated phase, in strongly-interacting systems of fermions or
hardcore bosons. There are numerous works devoted to this problem,
where it is studied, by approximate methods, in the framework of
relevant for experiments models. Unfortunately, due to tiny energy
differences involved, it is difficult to settle this problem by
means of approximate methods, which bias the calculated energies
with hardly controllable errors of various nature. We have studied
simple models, that by many physicists can be considered less
realistic, but which, in return, are amenable to a rigorous
analysis. Our models are described by extended Falicov--Kimball
Hamiltonians, with hoping particles being spinless fermions or
hardcore bosons. The ground-state phase diagrams of these models
have been constructed rigorously in the regime of strong coupling
and half-filling. In the both cases, of fermions and hardcore
bosons, we have found transitions from a crystalline chessboard
phase to the segregated phase via striped phases.

\ack V.D. is grateful to the University of Wroc{\l}aw, and
especially to the Institute of Theoretical Physics, for financial
support. This work was partially supported by Scientific Research
Grant {\textbf{2479/W/IFT}} (University of Wroc{\l}aw).

\section{Appendix}

In this section, at each point $p=(\omega,\varepsilon)$ we give
the sets $\{\alpha_{i}(p)\}$ of the zero-potential coefficients
$\alpha_{i}$ that we used to construct the four phase diagrams,
presented in the previous section.  Specifically, for each of the
four phase diagrams we define a partition, called
$\alpha$-partition, of the $(\omega,\varepsilon)$-plane (see the
tables displayed below) into sets that differ, in general, from
the domains ${\mathcal{S}}_{D}$. To each set of such a partition
we assign a set $\{(a_{i},b_{i},c_{i})\}$ of triplets of rational
numbers, such that the coefficients $\alpha_{i}(p)$ (with $p$ in
the considered set) are affine functions of the form $a_{i}\omega
+ b_{i}\varepsilon + c_{i}$. We have not been able to find one set
$\{(a_{i},b_{i},c_{i})\}$, that turns the fourth order potentials
into $m$-potentials in the whole $(\omega,\varepsilon)$-plane. We
have not also succeeded in assigning to each domain
${\mathcal{S}}_{D}$ an exactly one set $\{(a_{i},b_{i},c_{i})\}$.
The same remarks apply to the $(\omega,\delta)$-plane.
\begin{figure}[hbp]
\centering\includegraphics[height=0.35\textheight]{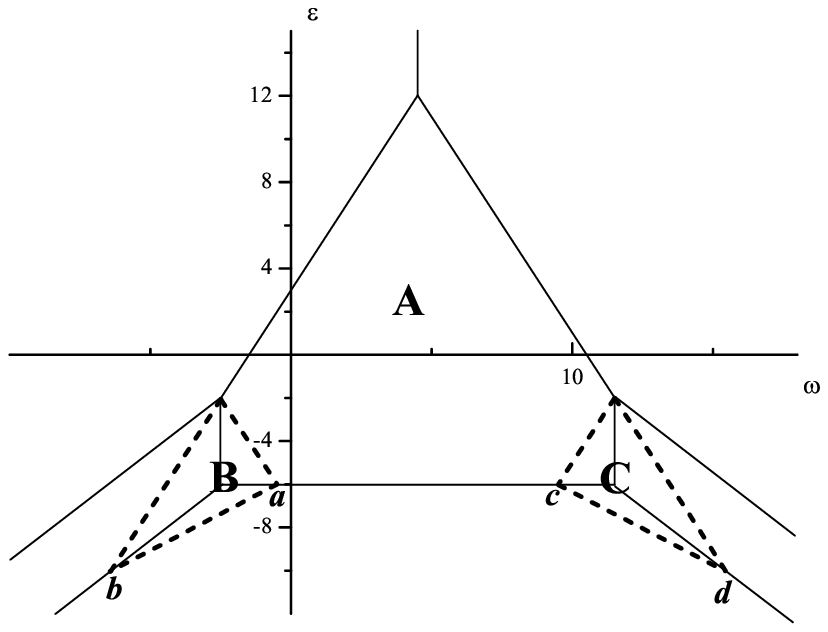}
\parbox[t]{\textwidth}
{\caption{{\small{The sets {\bf{A}}, {\bf{B}}, {\bf{C}},
whose boundaries are marked with dashed lines, used in Table~\ref{tb1}
to define the $\alpha$-partition of the fourth order phase diagram
in the case of fermions and for $\delta=0$.The dashed-line segments
are determined by their intersection points: ${\bf a}=(-1/2,-6)$,
${\bf b}=(-13/2,-10)$, ${\bf c}=(19/2,-6)$, ${\bf d}=(31/2,-10)$. The
coordinates of the remaining points can be read from
Fig.~\ref{t40f}.}}} \label{part1}} \vfill
\centering\includegraphics[height=0.35\textheight]{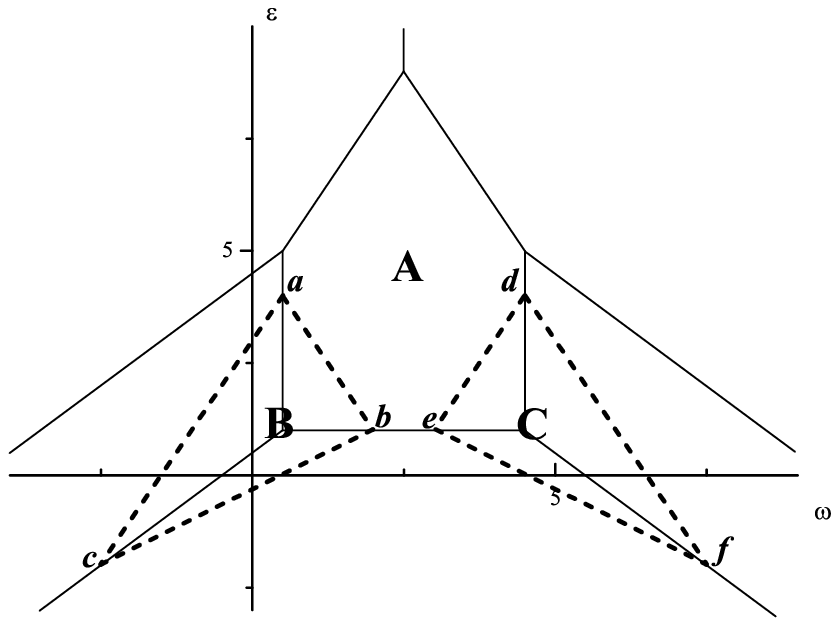}
\parbox[t]{\textwidth}
{\caption{{\small{The sets {\bf{A}}, {\bf{B}}, {\bf{C}}, whose boundaries
are marked with dashed lines, used in Table~\ref{tb2} to define the
$\alpha$-partition of the fourth order phase diagram in the case of
hardcore bosons and for $\delta=0$. The dashed-line segments are
determined by their intersection points: ${\bf a}=(1/2,4)$, ${\bf
b}=(2,1)$, ${\bf c}=(-5/2,-2)$, ${\bf d}=(9/2,4)$, ${\bf e}=
(3,1)$, ${\bf f} = (15/2,-2)$.}}} \label{part2}}
\end{figure}
\begin{table}[hbp]
\scriptsize
\begin{center}
\caption{The set of zero-potential coefficients $\{\alpha_{i}\}$ in
the case of fermions and for $\delta=0$. In the first column the
sets of the $\alpha$-partition are specified. For more comments see
the text in Appendix.} \label{tb1}
\begin{tabular}{|l|c|c|c|c|c|}
\hline & $\alpha_{1}$ & $\alpha_{2}$ & $\alpha_{3}$ & $\alpha_{4}$ &
$\alpha_{5}$
\\
\hline {\bf{A}} & $0$ & $0$ & $-\frac{\omega}{96}+\frac{3}{64}$ &
$0$ & $-\frac{1}{48}$
\\
\hline ${\mathcal{S}}_{dd}\cap${\bf{B}} & $0$ & $0$ & $-\frac{5
\omega}{192} -\frac{\varepsilon}{64} -\frac{3}{128}$ &
$\frac{\omega}{32} +\frac{\varepsilon}{32} +\frac{9}{64}$ &
$\frac{\omega}{64} +\frac{\varepsilon}{64} +\frac{19}{384}$
\\
\hline ${\mathcal{S}}_{v1}\cap${\bf{B}} & $0$ & $0$ &
$\frac{\omega}{192} -\frac{\varepsilon}{64} +\frac{7}{128}$ &
$-\frac{\omega}{32} +\frac{\varepsilon}{32} -\frac{1}{64}$ &
$-\frac{\omega}{64} +\frac{\varepsilon}{64} -\frac{11}{384}$
\\
\hline ${\mathcal{S}}_{v2}\cap${\bf{B}} & $0$ & $0$ & $-\frac{5
\omega}{192} +\frac{\varepsilon}{64} +\frac{21}{128}$ &
$\frac{\omega}{32} -\frac{\varepsilon}{32} -\frac{15}{64}$ &
$\frac{\omega}{64} -\frac{\varepsilon}{64} -\frac{53}{384}$
\\
\hline ${\mathcal{S}}_{dd}\cap${\bf{C}} & $0$ & $0$ & $-\frac{5
\omega}{192} +\frac{\varepsilon}{64} +\frac{33}{128}$ &
$-\frac{\omega}{32} +\frac{\varepsilon}{32} +\frac{27}{64}$ &
$-\frac{\omega}{64} +\frac{\varepsilon}{64} +\frac{73}{384}$
\\
\hline ${\mathcal{S}}_{v2}\cap${\bf{C}} & $0$ & $0$ & $-\frac{5
\omega}{192} -\frac{\varepsilon}{64} +\frac{9}{128}$ &
$-\frac{\omega}{32} -\frac{\varepsilon}{32} +\frac{3}{64}$ &
$-\frac{\omega}{64} -\frac{\varepsilon}{64} +\frac{1}{384}$
\\
\hline ${\mathcal{S}}_{d1}\cap${\bf{C}} & $0$ & $0$ &
$\frac{\omega}{192} +\frac{\varepsilon}{64} -\frac{13}{128}$ &
$\frac{\omega}{32} +\frac{\varepsilon}{32} -\frac{19}{64}$ &
$\frac{\omega}{64} +\frac{\varepsilon}{64} -\frac{65}{384}$
\\
\hline
\end{tabular}
\end{center}
\end{table}
\begin{table}[ht]
\scriptsize
\begin{center}
\caption{The set of zero-potential coefficients $\{\alpha_{i}\}$ in
the case of hardcore bosons and for $\delta=0$. In the first column
the sets of the $\alpha$-partition are specified. For more comments
see the text in Appendix.} \label{tb2}
\begin{tabular}{|l|c|c|c|c|c|}
\hline & $\alpha_{1}$ & $\alpha_{2}$ & $\alpha_{3}$ & $\alpha_{4}$ &
$\alpha_{5}$
\\
\hline {\bf{A}} & $0$ & $0$ & $-\frac{\omega}{96}+\frac{5}{192}$ &
$0$ & $-\frac{1}{48}$
\\
\hline ${\mathcal{S}}_{pcb}\cap${\bf{B}} & $0$ & $0$ &
$-\frac{\omega}{24} -\frac{\varepsilon}{64} +\frac{23}{192}$ &
$\frac{\omega}{16} +\frac{7 \varepsilon}{192} -\frac{37}{192}$ &
$\frac{\omega}{32} +\frac{5 \varepsilon}{256} -\frac{91}{768}$
\\
\hline ${\mathcal{S}}_{v1}\cap${\bf{B}} & $0$ & $0$ &
$\frac{\omega}{192} -\frac{\varepsilon}{64} +\frac{37}{384}$ &
$-\frac{7 \omega}{192} +\frac{7 \varepsilon}{192} -\frac{55}{384}$ &
$-\frac{5 \omega}{256} +\frac{5 \varepsilon}{256} -\frac{143}{1536}$
\\
\hline ${\mathcal{S}}_{v2}\cap${\bf{B}} & $0$ & $0$ &
$-\frac{\omega}{24} +\frac{\varepsilon}{32} +\frac{7}{96}$ &
$\frac{\omega}{16} -\frac{\varepsilon}{16} -\frac{3}{32}$ &
$\frac{\omega}{32} -\frac{\varepsilon}{32} -\frac{13}{192}$
\\
\hline ${\mathcal{S}}_{pcb}\cap${\bf{C}} & $0$ & $0$ &
$-\frac{\omega}{24} +\frac{\varepsilon}{64} +\frac{17}{192}$ &
$-\frac{\omega}{16} +\frac{7 \varepsilon}{192} +\frac{23}{192}$ &
$-\frac{\omega}{32} +\frac{5 \varepsilon}{256} +\frac{29}{768}$
\\
\hline ${\mathcal{S}}_{v2}\cap${\bf{C}} & $0$ & $0$ &
$-\frac{\omega}{24} -\frac{\varepsilon}{32} +\frac{13}{96}$ &
$-\frac{\omega}{16} -\frac{\varepsilon}{16} +\frac{7}{32}$ &
$-\frac{\omega}{32} -\frac{\varepsilon}{32} +\frac{17}{192}$
\\
\hline ${\mathcal{S}}_{d1}\cap${\bf{C}} & $0$ & $0$ &
$\frac{\omega}{192} +\frac{\varepsilon}{64} -\frac{47}{384}$ &
$\frac{7 \omega}{192} +\frac{7 \varepsilon}{192} -\frac{125}{384}$ &
$\frac{5 \omega}{256} +\frac{5 \varepsilon}{256} -\frac{293}{1536}$
\\
\hline
\end{tabular}
\end{center}
\end{table}
\begin{figure}[ht]
\centering\includegraphics[height=0.35\textheight]{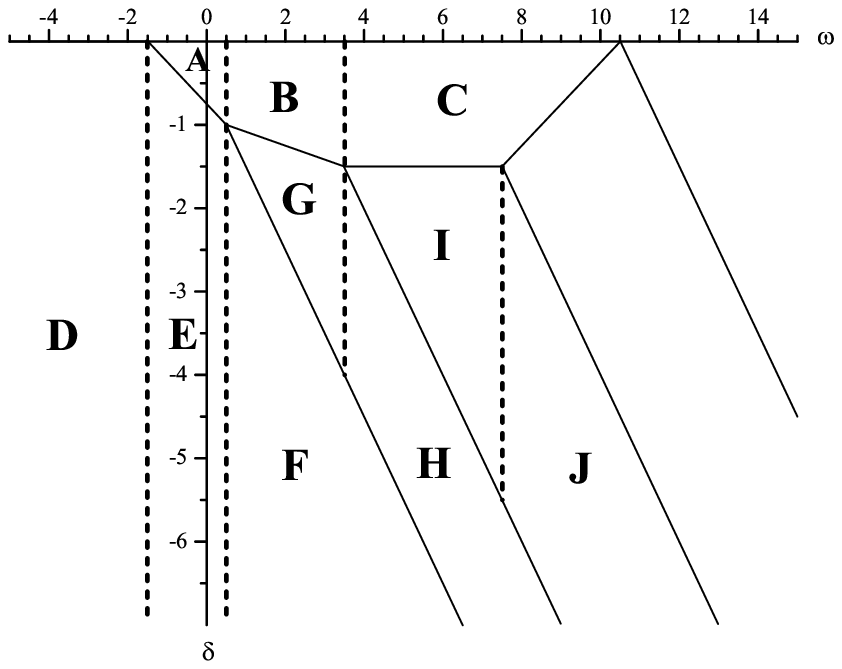}
\parbox[t]{\textwidth}
{\caption{{\small{The sets {\bf A,B},\ldots, whose boundaries are
marked with dashed and continuous lines (the continuous lines are
the boundaries of the phase diagram), used in Table~\ref{tb3} to
define the $\alpha$-partition of the fourth order phase diagram in
the case of fermions and for $\varepsilon=0$.}}} \label{part3}}
\vfill \centering\includegraphics[height=0.35\textheight]{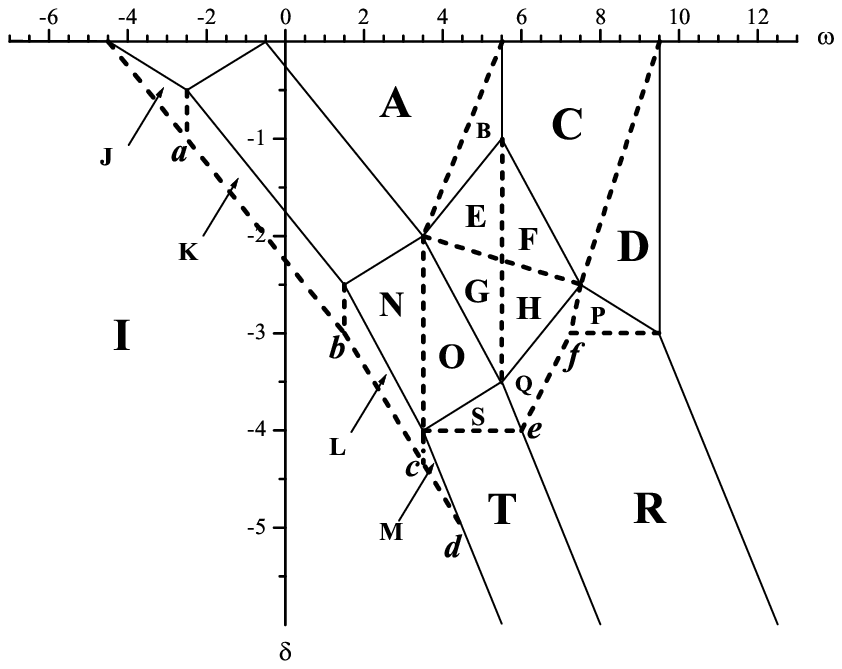}
\parbox[t]{\textwidth}
{\caption{{\small{The sets {\bf A,B},\ldots, whose boundaries are
marked with dashed and continuous lines (the continuous lines are
the boundaries of the phase diagram), used in Table~\ref{tb4} to
define the $\alpha$-partition of the fourth order phase diagram in
the case of hardcore bosons and for $\varepsilon=0$. The dashed-line
segments are determined by their intersection points: ${\bf a} =
(-5/2,-1)$, ${\bf b} = (3/2,-3)$, ${\bf c} = (7/2,-13/3)$, ${\bf d}
= (9/2,-5)$, ${\bf e} = (6,-4)$, ${\bf f} = (29/4,-3)$.}}}
\label{part4}}
\end{figure}
\begin{table}[t]
\scriptsize
\begin{center}
\caption{The set of zero-potential coefficients $\{\alpha_{i}\}$ in
the case of fermions and for $\varepsilon=0$. In the first column
the sets of the $\alpha$-partition are specified. For more comments
see the text in Appendix.} \label{tb3}
\begin{tabular}{|l|c|c|c|c|c|}
\hline & $\alpha_{1}$ & $\alpha_{2}$ & $\alpha_{3}$ & $\alpha_{4}$ &
$\alpha_{5}$
\\
\hline ${\mathcal{S}}_{-}\cap${\bf{D}} & $-\frac{201 \delta}{3200}$
& $-\frac{\delta}{8}$ & $-\frac{\omega}{96} +\frac{\delta}{6400}
+\frac{3}{64}$ & $0$ & $-\frac{1}{48}$
\\
\hline ${\mathcal{S}}_{-}\cap${\bf{E}} & $\frac{\omega}{32}
-\frac{\delta}{16} +\frac{3}{64}$ & $-\frac{\delta}{8}$ &
$-\frac{\omega}{96} +\frac{3}{64}$ & $0$ & $-\frac{1}{48}$
\\
\hline ${\mathcal{S}}_{-}\cap${\bf{F}} & $\frac{\omega}{48}
-\frac{\delta}{16} +\frac{5}{96}$ & $-\frac{\delta}{8}$ &
$\frac{1}{24}$ & $0$ & $-\frac{1}{48}$
\\
\hline ${\mathcal{S}}_{1}\cap${\bf{G}} & $\frac{11
\omega}{960}-\frac{9 \delta}{160}+ \frac{121}{1920}$ &
$-\frac{\delta}{8}$ & $-\frac{7 \omega}{5760}+ \frac{\delta}{320}+
\frac{403}{11520}$ &
$-\frac{\omega}{180}-\frac{\delta}{80}-\frac{11}{360}$ &
$-\frac{\omega}{144}+\frac{1}{288}$
\\
\hline ${\mathcal{S}}_{1}\cap${\bf{H}} & $-\frac{3
\omega}{160}-\frac{13 \delta}{160}+ \frac{11}{160}$ &
$\frac{\omega}{40}-\frac{\delta}{10}+\frac{1}{80}$ & $-\frac{13
\omega}{960}-\frac{\delta}{320}+\frac{17}{320}$ &
$-\frac{\omega}{40}-\frac{\delta}{40}-\frac{1}{80}$ &
$-\frac{1}{48}$
\\
\hline ${\mathcal{S}}_{2}\cap${\bf{I}} & $-\frac{9
\delta}{128}+\frac{21}{256}$ & $-\frac{5 \delta}{64}+\frac{9}{128}$
& $-\frac{\omega}{96}+\frac{\delta}{128}+\frac{19}{256}$ & $-\frac{3
\delta}{128}-\frac{17}{256}$ & $\frac{\omega}{512}-\frac{13}{3072}$
\\
\hline ${\mathcal{S}}_{2}\cap${\bf{J}} & $\frac{\omega}{128}
-\frac{7 \delta}{128}+\frac{3}{64}$ & $-\frac{\omega}{64}-\frac{9
\delta}{64}+\frac{3}{32}$ & $-\frac{\omega}{96}+\frac{1}{16}$ &
$\frac{\omega}{128}+\frac{\delta}{128}-\frac{5}{64}$ &
$\frac{\omega}{128}+\frac{\delta}{128}-\frac{7}{192}$
\\
\hline ${\mathcal{S}}_{3}$ & $-\frac{7 \omega}{384} -\frac{17
\delta}{192}+ \frac{49}{256}$ & $-\frac{\delta}{8}$ & $-\frac{29
\omega}{2304}- \frac{7 \delta}{1152}+\frac{107}{1536}$ &
$\frac{\omega}{144}+ \frac{\delta}{144}-\frac{7}{96}$ &
$-\frac{\omega}{144}-\frac{\delta}{144}+\frac{5}{96}$
\\
\hline ${\mathcal{S}}_{cb}$ & $-\frac{359 \delta}{5728}$ &
$-\frac{\delta}{8}$ & $-\frac{173 \omega}{17184}+ \frac{3
\delta}{11456}+ \frac{495}{11456}$ & $0$ & $-\frac{1}{48}$
\\
\hline ${\mathcal{S}}_{dd}\cap${\bf{A}} & $-\frac{\delta}{8}$ &
$-\frac{\delta}{8}$ & $-\frac{\omega}{96}+ \frac{\delta}{96}+
\frac{3}{64}$ & $\frac{\delta}{48}$ &
$-\frac{\delta}{48}-\frac{1}{48}$
\\
\hline ${\mathcal{S}}_{dd}\cap${\bf{B}} & $-\frac{\delta}{8}$ &
$-\frac{\delta}{8}$ & $\frac{\delta}{96}+ \frac{1}{24}$ &
$\frac{\delta}{48}$ & $-\frac{\delta}{48}-\frac{1}{48}$
\\
\hline ${\mathcal{S}}_{dd}\cap${\bf{C}} & $-\frac{\delta}{8}$ &
$-\frac{\delta}{8}$ & $- \frac{\omega}{96}- \frac{\delta}{96}+
\frac{3}{64}$ & $\frac{\delta}{48}$ &
$-\frac{\delta}{48}-\frac{1}{48}$
\\
\hline
\end{tabular}
\end{center}
\end{table}
\begin{table}[t]
\scriptsize
\begin{center}
\caption{The set of zero-potential coefficients $\{\alpha_{i}\}$ in
the case of hardcore bosons and for $\varepsilon=0$. In the first
column the sets of the $\alpha$-partition are specified. The cases,
where the set ${\mathcal{S}}_{D|T}$ is a proper subset of
${\mathcal{S}}_{TD}(p)$ are marked by the asterisk. For more
comments see the text in Appendix.} \label{tb4}
\begin{tabular}{|l|c|c|c|c|c|}
\hline & $\alpha_{1}$ & $\alpha_{2}$ & $\alpha_{3}$ & $\alpha_{4}$ &
$\alpha_{5}$
\\
\hline ${\mathcal{S}}_{-}\cap${\bf{I}} & $-\frac{\delta}{16}$ &
$-\frac{\delta}{8}$ & $-\frac{\omega}{96}-\frac{1}{192}$ & $0$ &
$-\frac{1}{48}$
\\
\hline ${\mathcal{S}}_{-}\cap${\bf{J}} & $\frac{\omega}{32}
-\frac{\delta}{16} +\frac{9}{64}$ & $-\frac{\delta}{8}$ & $\frac{7
\omega}{192} +\frac{3 \delta}{32} +\frac{91}{384}$ & $-\frac{3
\omega}{64} -\frac{3 \delta}{32} -\frac{27}{128}$ & $-\frac{1}{48}$
\\
\hline ${\mathcal{S}}_{-}\cap${\bf{K}} & $-\frac{3 \omega}{128}
-\frac{\delta}{16} -\frac{7}{256}$ & $\frac{3 \omega}{64}
-\frac{\delta}{8} +\frac{23}{128}$ & $\frac{\omega}{48}
+\frac{\delta}{16} +\frac{19}{96}$ & $-\frac{\omega}{32} -\frac{3
\delta}{64} -\frac{21}{128}$ & $-\frac{15 \omega}{512} -\frac{7
\delta}{256} -\frac{331}{3072}$
\\
\hline ${\mathcal{S}}_{-}\cap${\bf{L}} & $\frac{\omega}{16}
-\frac{\delta}{16} -\frac{5}{32}$ & $-\frac{\omega}{8}
-\frac{\delta}{8} +\frac{7}{16}$ & $\frac{11 \omega}{384} +\frac{3
\delta}{32} +\frac{203}{768}$ & $\frac{21 \omega}{128} +\frac{5
\delta}{32} +\frac{13}{256}$ & $-\frac{41 \omega}{512} -\frac{19
\delta}{128} -\frac{1027}{3072}$
\\
\hline ${\mathcal{S}}_{-}\cap${\bf{M}} & $-\frac{\delta}{16}
+\frac{1}{16}$ & $-\frac{\omega}{32} -\frac{5 \delta}{32}
-\frac{1}{64}$ & $-\frac{23 \omega}{768} -\frac{5 \delta}{256}
+\frac{25}{1536}$ & $\frac{7 \omega}{128} +\frac{7 \delta}{128}
+\frac{7}{256}$ & $-\frac{\omega}{128} -\frac{\delta}{128}
-\frac{19}{768}$
\\
\hline ${\mathcal{S}}_{1}\cap${\bf{S}} & $-\frac{9 \omega}{160}
-\frac{11 \delta}{40} -\frac{189}{320}$ & $\frac{\omega}{20}
+\frac{\delta}{20} +\frac{21}{40}$ & $-\frac{5 \omega}{192}
-\frac{\delta}{16} -\frac{65}{384}$ & $-\frac{3 \omega}{80}
-\frac{\delta}{10} -\frac{43}{160}$ & $-\frac{\omega}{80} -\frac{3
\delta}{40} -\frac{133}{480}$
\\
\hline ${\mathcal{S}}_{1}\cap${\bf{T}} & $-\frac{\omega}{40}
-\frac{7 \delta}{80} +\frac{1}{20}$ & $\frac{\omega}{40}
-\frac{\delta}{10} +\frac{1}{80}$ & $-\frac{\omega}{60}
-\frac{\delta}{160} +\frac{11}{480}$ & $-\frac{\omega}{40}
-\frac{\delta}{40} -\frac{1}{80}$ & $-\frac{1}{48}$
\\
\hline ${\mathcal{S}}_{3}\cap${\bf{P}} & $-\frac{\omega}{192}
-\frac{\delta}{12} -\frac{5}{384}$ & $-\frac{\delta}{8}$ &
$-\frac{\omega}{288} -\frac{5 \delta}{144} -\frac{83}{576}$ &
$\frac{\omega}{192} -\frac{\delta}{24} -\frac{67}{384}$ &
$-\frac{\omega}{288} -\frac{\delta}{72} -\frac{17}{576}$
\\
\hline ${\mathcal{S}}_{3}\cap${\bf{Q}} & $\frac{25 \omega}{192}
-\frac{\delta}{6} -\frac{475}{384}$ & $-\frac{\omega}{8}
-\frac{\delta}{16} +\frac{35}{32}$ & $\frac{3 \omega}{128} -\frac{7
\delta}{192} -\frac{269}{768}$ & $\frac{\omega}{16}
-\frac{\delta}{32} -\frac{37}{64}$ & $\frac{5 \omega}{96}
-\frac{\delta}{24} -\frac{33}{64}$
\\
\hline ${\mathcal{S}}_{3}\cap${\bf{R}} & $-\frac{\delta}{16}$ &
$-\frac{\omega}{72} -\frac{5 \delta}{36} +\frac{13}{144}$ &
$-\frac{\omega}{144} +\frac{\delta}{288} +\frac{1}{288}$ &
$\frac{\omega}{72} +\frac{\delta}{72} -\frac{13}{144}$ &
$-\frac{1}{48}$
\\
\hline ${\mathcal{S}}_{cb}$ & $-\frac{\delta}{16}$ &
$-\frac{\delta}{8}$ & $-\frac{\omega}{96} +\frac{5}{192}$ & $0$ &
$-\frac{1}{48}$
\\
\hline ${\mathcal{S}}_{v1}$ & $-\frac{3 \delta}{16}$ &
$-\frac{\delta}{8}$ & $\frac{\omega}{192} +\frac{37}{384}$ &
$-\frac{\omega}{32} -\frac{\delta}{16} -\frac{9}{64}$ &
$-\frac{\omega}{64} -\frac{\delta}{8} -\frac{35}{384}$
\\
\hline ${\mathcal{S}}_{v2}\cap${\bf{A}} & $-\frac{\delta}{32}$ &
$-\frac{3 \delta}{16}$ & $-\frac{\omega}{32} -\frac{\delta}{96}
+\frac{5}{64}$ & $-\frac{1}{8}$ & $-\frac{1}{12}$
\\
\hline ${\mathcal{S}}_{v2}\cap${\bf{B}} $^{*}$ & $ \frac{\omega}{32}
-\frac{\delta}{16} -\frac{11}{64}$ & $-\frac{\omega}{16}
-\frac{\delta}{8} +\frac{11}{32}$ & $-\frac{\omega}{24}
+\frac{13}{96}$ & $-\frac{1}{8}$ & $-\frac{1}{12}$
\\
\hline ${\mathcal{S}}_{v3}$ & $-\frac{\omega}{192}
-\frac{\delta}{24} -\frac{1}{384}$ & $-\frac{\omega}{32}
-\frac{\delta}{4} -\frac{1}{64}$ & $-\frac{7 \omega}{192}
-\frac{\delta}{48} +\frac{29}{384}$ & $-\frac{\omega}{96}
-\frac{\delta}{48} -\frac{25}{192}$ & $-\frac{1}{12}$
\\
\hline ${\mathcal{S}}_{d1}\cap${\bf{C}} & $-\frac{\delta}{16}$ &
$-\frac{\delta}{8}$ & $\frac{\omega}{192} -\frac{47}{384}$ &
$\frac{3 \omega}{64} -\frac{49}{128}$ & $\frac{\omega}{32}
-\frac{49}{192}$
\\
\hline ${\mathcal{S}}_{d1}\cap${\bf{D}} & $-\frac{\delta}{16}$ &
$-\frac{\delta}{8}$ & $\frac{\omega}{192} -\frac{47}{384}$ &
$\frac{\omega}{64} -\frac{19}{128}$ & $-\frac{1}{48}$
\\
\hline ${\mathcal{S}}_{d2}\cap${\bf{N}} & $\frac{\omega}{64}
-\frac{\delta}{8} -\frac{31}{128}$ & $-\frac{5 \omega}{64}
-\frac{\delta}{16} +\frac{67}{128}$ & $-\frac{\omega}{24}
+\frac{13}{96}$ & $-\frac{\delta}{16} -\frac{1}{4}$ &
$\frac{\omega}{128} -\frac{\delta}{32} -\frac{133}{768}$
\\
\hline ${\mathcal{S}}_{d2}\cap${\bf{O}} & $-\frac{3 \omega}{32}
-\frac{\delta}{8} +\frac{9}{64}$ & $\frac{5 \omega}{64}
-\frac{\delta}{16} -\frac{3}{128}$ & $-\frac{\omega}{24}
+\frac{13}{96}$ & $-\frac{3 \omega}{64} -\frac{\delta}{16}
-\frac{11}{128}$ & $-\frac{3 \omega}{128} -\frac{\delta}{32}
-\frac{49}{768}$
\\
\hline ${\mathcal{S}}_{d3}\cap${\bf{E}} $^{*}$ & $\frac{\omega}{32}
-\frac{\delta}{16} -\frac{11}{64}$ & $-\frac{\omega}{20} -\frac{3
\delta}{20} +\frac{1}{4}$ & $-\frac{\omega}{60} -\frac{\delta}{20}
-\frac{5}{96}$ & $\frac{\omega}{32} -\frac{\delta}{16}
-\frac{23}{64}$ & $\frac{3 \omega}{160} -\frac{3 \delta}{80}
-\frac{43}{192}$
\\
\hline ${\mathcal{S}}_{d3}\cap${\bf{F}} $^{*}$ &
$-\frac{\delta}{16}$ & $-\frac{3 \omega}{160} -\frac{3 \delta}{20}
+\frac{5}{64}$ & $-\frac{31 \omega}{960} -\frac{\delta}{20}
+\frac{13}{384}$ & $-\frac{\delta}{16} -\frac{3}{16}$ &
$\frac{\omega}{320} -\frac{3 \delta}{80} -\frac{53}{384}$

\\
\hline ${\mathcal{S}}_{d3}\cap${\bf{G}} $^{*}$ & $\frac{3
\omega}{64} +\frac{\delta}{16} +\frac{3}{128}$ & $-\frac{\omega}{16}
-\frac{\delta}{4} +\frac{3}{32}$ & $-\frac{\omega}{240}
+\frac{\delta}{20} +\frac{5}{48}$ & $\frac{3 \omega}{64}
+\frac{\delta}{16} -\frac{21}{128}$ & $\frac{9 \omega}{320} +\frac{3
\delta}{80} -\frac{41}{384}$
\\
\hline ${\mathcal{S}}_{d3}\cap${\bf{H}} $^{*}$ & $\frac{\omega}{64}
+\frac{\delta}{16} +\frac{25}{128}$ & $-\frac{\omega}{32}
-\frac{\delta}{4} -\frac{5}{64}$ & $-\frac{19 \omega}{960}
+\frac{\delta}{20} +\frac{73}{384}$ & $\frac{\omega}{64}
+\frac{\delta}{16} +\frac{1}{128}$ & $\frac{\omega}{80} +\frac{3
\delta}{80} -\frac{1}{48}$
\\
\hline
\end{tabular}
\end{center}
\end{table}

\end{document}